\documentclass[twocolumn]{aastex62}

\usepackage{nameref} 

\newcommand{\Teff}{$T_{\rm{eff}}$}

\newcommand{\Msun}{$M_{\odot}$}

\newcommand{\newrev}{ }

\newcommand{\sect}{Sect.\,}

\usepackage{xcolor}
\usepackage{graphicx}	
\usepackage{amsmath}	
\usepackage{multirow}
\usepackage{appendix}
\usepackage{chngcntr} 

\usepackage{longtable} 
\usepackage{subfigure}

\def\arcsec{\hbox{$^{\hbox{\rlap{\hbox{\lower4pt\hbox{$\,\prime\prime$}}}\hbox{$\frown$}}}$}}

\graphicspath{{./}{figures/}}

\shorttitle{}
\shortauthors{}

\usepackage{lineno}
\begin{document}


\title{A comprehensive Gaia spectroscopic study of stars in the Sco-Cen complex:
 star formation history and disk lifetime}

\correspondingauthor{Min Fang}
\email{mfang@pmo.ac.cn}

\author{Min Fang}
\affiliation{Purple Mountain Observatory, Chinese Academy of Sciences, 10 Yuanhua Road, Nanjing 210023, People's Republic of China}
\affiliation{University of Science and Technology of China, Hefei 230026, People's Republic of China}

\author{Gregory J. Herczeg}
\affiliation{Kavli Institute for Astronomy and Astrophysics, Peking University, No.5 Yiheyuan Road, Haidian District, Beijing 100871, People's Republic of China}
\affiliation{Department of Astronomy, Peking University, No.5 Yiheyuan Road, Haidian District, Beijing 100871, People's Republic of China}

\begin{abstract}
We re-evaluate the star formation history of the nearby Sco-Cen OB Association with a comprehensive analysis of Gaia XP spectra of more than 7,800 potential members.
New spectral classifications are obtained by fitting individual XP spectra with templates derived from empirical spectra of young stars.  
 Combining these spectral classifications in this work and in the literature with archival photometry leads to estimates of V-band extinction and stellar luminosities for a total of 8,846 sources.
 Employing SPOTS models with spot coverages of 0.34 and 0.51 for K and M-type stars harmonizes age estimates between K/M-type and F/G-type stars, with ages older than are obtained for low-mass stars from standard evolutionary models.  These older ages lead to a disk lifetime that is approximately two times longer than reported in previous literature.
Our re-evaluation of the star formation history with these revised age estimates uncovers evidence of underlying substructures within the Sco-Cen complex. 
 \end{abstract}


\section{Introduction} 
\label{sec:intr}
The Scorpius-Centaurus (Sco-Cen) complex, one of the nearest and most extensively studied OB associations to the Sun, has been a cornerstone for understanding the
structure, dynamics, and history of star-forming regions. 
Kinematic and spectroscopic analyses of members identified with Hipparcos data revealed a complex sub-structure composed of several distinct subgroups, including Upper Scorpius (US), Upper Centaurus Lupus (UCL), and Lower Centaurus Crux (LCC) \citep{1999AJ....117..354D}. A significant body of research  concentrated on the membership and age determinations of the Sco-Cen  members\citep{1999AJ....117.2381P,2006AJ....131.3016S,2008ApJ...688..377S,2012ApJ...746..154P,2016A&A...593A..99F,2017ApJ...842..123F,2023ApJ...954..134K} have provided more accurate estimates, indicating that the US subgroup is the youngest, with an age of approximately 5-10 Myr \citep{1999AJ....117.2381P,2012ApJ...746..154P,2016A&A...593A..99F}, while UCL and LCC are older, around 16 and 17 Myr, respectively \citep{2012ApJ...746..154P}.

The high-precision astrometric data from the Gaia mission has significantly advanced our understanding of the Sco-Cen complex. The improvements in kinematic properties of stars has led to a more complete census of the association,  with thousands of newly identified members, and refined definitions of the substructures in the complex\citep{2019A&A...623A.112D,2020AJ....160...44L,2022AJ....163...24L,2023ApJ...954..134K}.

The comprehensive investigation of the Sco-Cen complex continues to enhance our knowledge of the physical and dynamical characteristics of nearby young stellar populations, serving as a benchmark for comparative studies of star-forming regions throughout the Galaxy. In this work, we use Gaia XP spectra, combined with other observations, to study previously identified members in the Sco-Cen complex in order to refine our understanding of the star formation process in this region.

\begin{figure}
\begin{center}
\includegraphics[width=1.0\columnwidth]{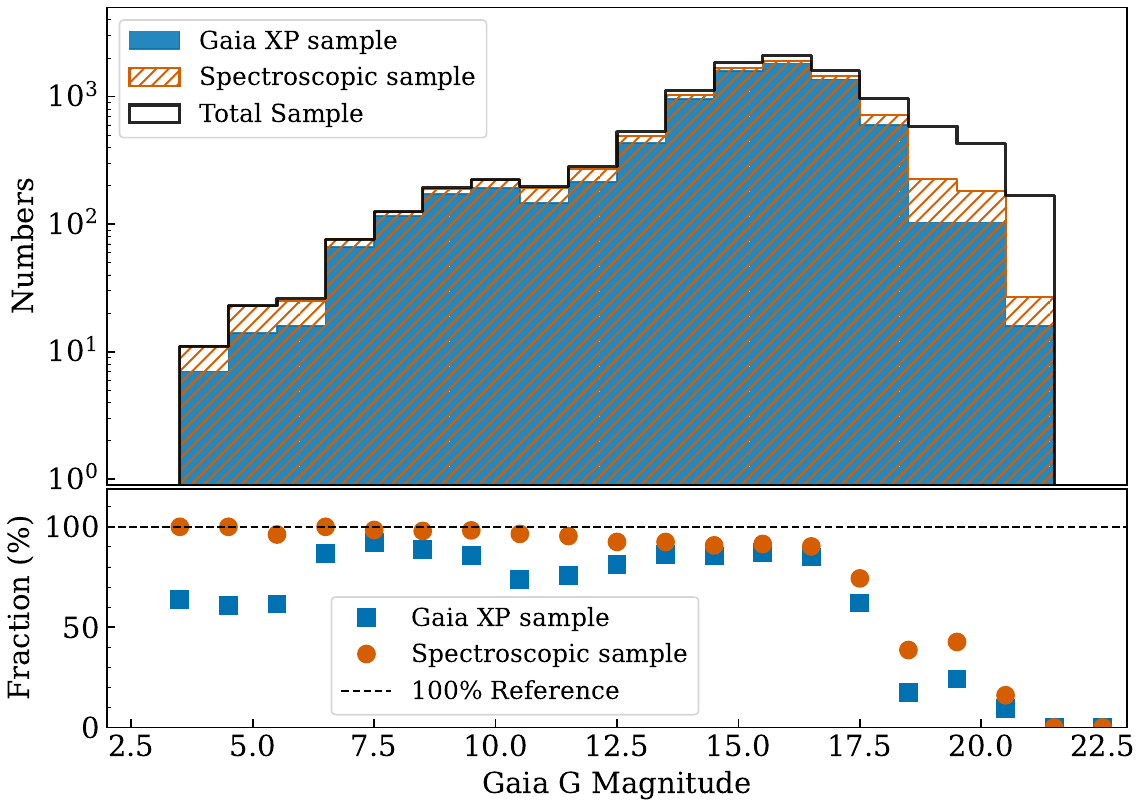}
\caption{Upper panel: Histograms depicting the Gaia G magnitude distribution for the candidate members of the Sco-Cen complex (black open histograms), the ones with the available Gaia XP spectra(blue-color filled histograms), and all the spectroscopic sample studied in this work (orange-line filled histograms). lower panel: Fractions of  sources with the Gaia XP spectra (blue filled squares) and all the spectroscopic ones (orange filled circles) among  all the candidate members of the Sco-Cen complex across individual Gaia G magnitude interval.}\label{Fig:spsample}
\end{center}
\end{figure}

\section{Data and analysis}
\label{sec:Data}

\subsection{data}
\subsubsection{Sample}
\label{sec:sample}

\begin{figure*}
\begin{center}
\includegraphics[width=2.0\columnwidth]{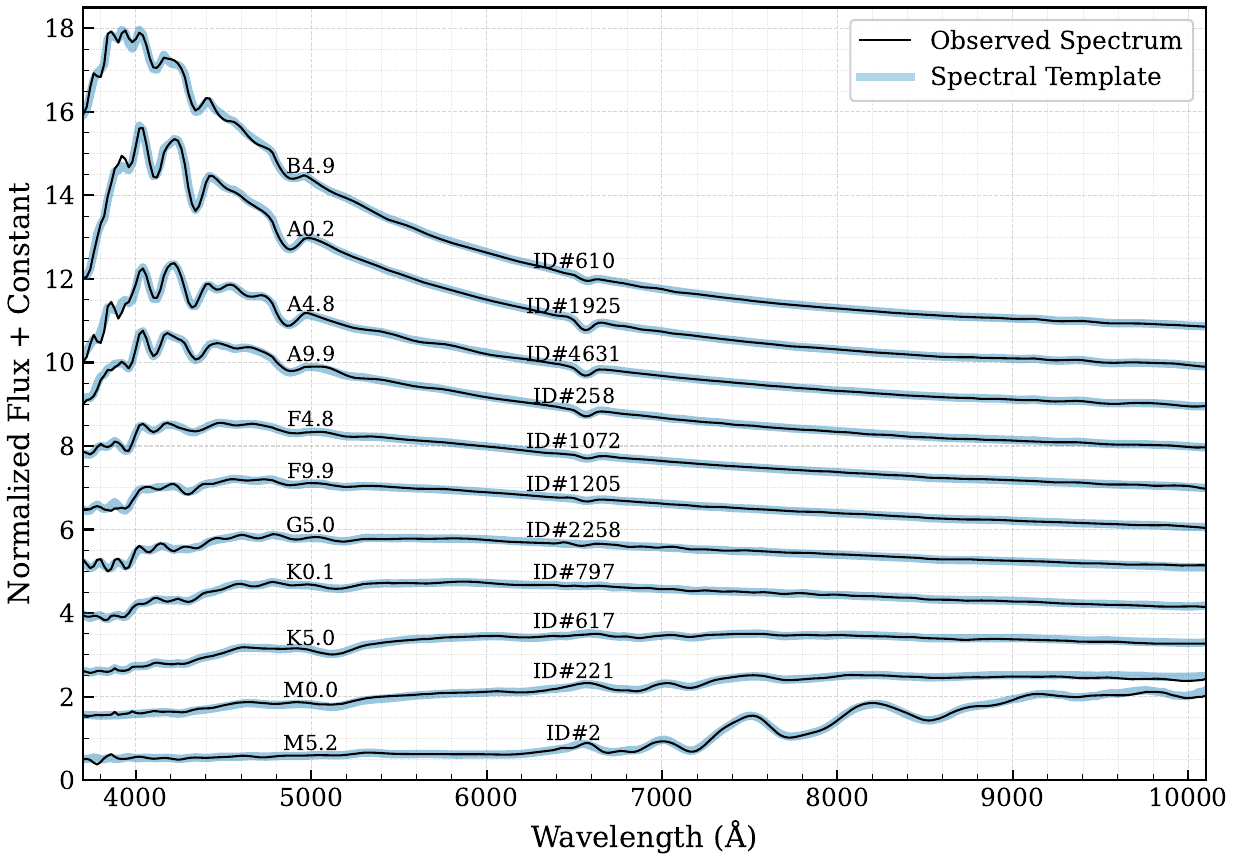}
\caption{
Example Gaia XP spectra (black thin lines) of 11 candidate members with their best-fit spectral templates ({\newrev blue} thick lines).
 }\label{Fig:fitexample}
\end{center}
\end{figure*}

The candidate members of the Sco-Cen Complex are collected from \cite{2022AJ....163...24L}. In that work, \cite{2022AJ....163...24L} used high-precision photometric and astrometric data from Gaia EDR3 to conduct a survey of stellar populations within the Sco-Cen Complex and identified 10,509 candidate members with parallax uncertainties less than 1 mas. These sources are distributed within the regions of Upper Sco, UCL/LCC, the V1062 Sco group, Ophiuchus, and Lupus. Figure~\ref{Fig:spsample} shows the Gaia G magnitude distribution for the candidate members of the Sco-Cen complex\footnote{Eight sources without the Gaia G magnitude in Gaia DR3 are not shown in Figure~\ref{Fig:spsample}.}. The distribution peaks at G=15.5, which corresponds to a 0.3\,\Msun\ pre-main sequence star  at an age of 10-20 Myr in the Sco-Cen complex according to the PARSEC models \citep{2012MNRAS.427..127B,2014MNRAS.444.2525C}.

\subsubsection{Gaia XP spectra}
\label{sec:gaiadr3}

Of the 10,509 candidate members, 7,923 have Gaia XP spectra in Gaia Data Release 3 (DR3). We refine the flux calibration of these XP spectra using the python code provided in \cite{2024ApJS..271...13H}. 
The corrected Gaia XP spectra show consistency with the MILES and LEMONY libraries within 2\% for wavelengths 336–400 nm and 1\% for longer wavelengths  \citep{2024ApJS..271...13H}. Figure~\ref{Fig:spsample} shows the numbers and the proportion of sources with the published Gaia XP spectra in Gaia DR3 across individual Gaia G magnitude interval. Among the member candidates with Gaia G-band magnitudes ranging from 6.5 to 16.5~mag, over 85\% of them have Gaia XP spectra. 
Example Gaia~XP spectra of the candidate members are shown in Figure~\ref{Fig:fitexample}.

\subsubsection{Photometric data}
\label{sec:photometry}

In order to construct the spectral energy distribution (SED) of each source, we used optical  photometry in  Pan-STARRS,
the $g$, $r$, $i$, $z$, and $y$ bands from 
the Gaia synthetic photometry catalogue \citep{2023A&A...674A..33G}, optical photometry in $g$, $r$, $i$, $z$ from SkyMapper Southern Sky Survey \citep[Data Release 4, ][]{2024arXiv240202015O},
$G$, $B_P$, and $R_P$ bands from Gaia Data Release 3 \citep{2023A&A...674A...1G}, near-infrared photometry in the $J$, $H$, and $K_S$ bands from the Two-Micron All Sky Survey \citep[2MASS, ][]{2006AJ....131.1163S}, infrared photometry in W1, W2, W3, and W4 from AllWISE Data Release \citep{2013wise.rept....1C}, and infrared photometry in 3.6\,$\mu$m, 4.5\,$\mu$m, 5.8\,$\mu$m, 8.0\,$\mu$m, and 24\,$\mu$m from the Spitzer Enhanced Imaging Products \citep[SEIP;][]{2012AAS...21942806T}

\subsection{Analysis}
\subsubsection{Spectral Classification}\label{sect:class}

We classify the Gaia XP spectra of individual sources by fitting their spectra with the Gaia~XP  spectral templates with spectral types ranging from B2 to M8. The templates are constructed using the observed Gaia XP spectra of a sample of young stars with known spectral types in the literature (see details in Appendix~\ref{Appen:template}). The spectral fitting includes two free parameters, spectral type (SPT) and visual extinction ($A_{\rm V}$). The extinction law used to redden the spectral templates is from \citet{1989ApJ...345..245C} with  a total to selective extinction value $R_{\rm V}=$3.1, typical of interstellar medium dust. Our process is as follows:

\begin{figure}
\begin{center}
\includegraphics[width=1.0\columnwidth]{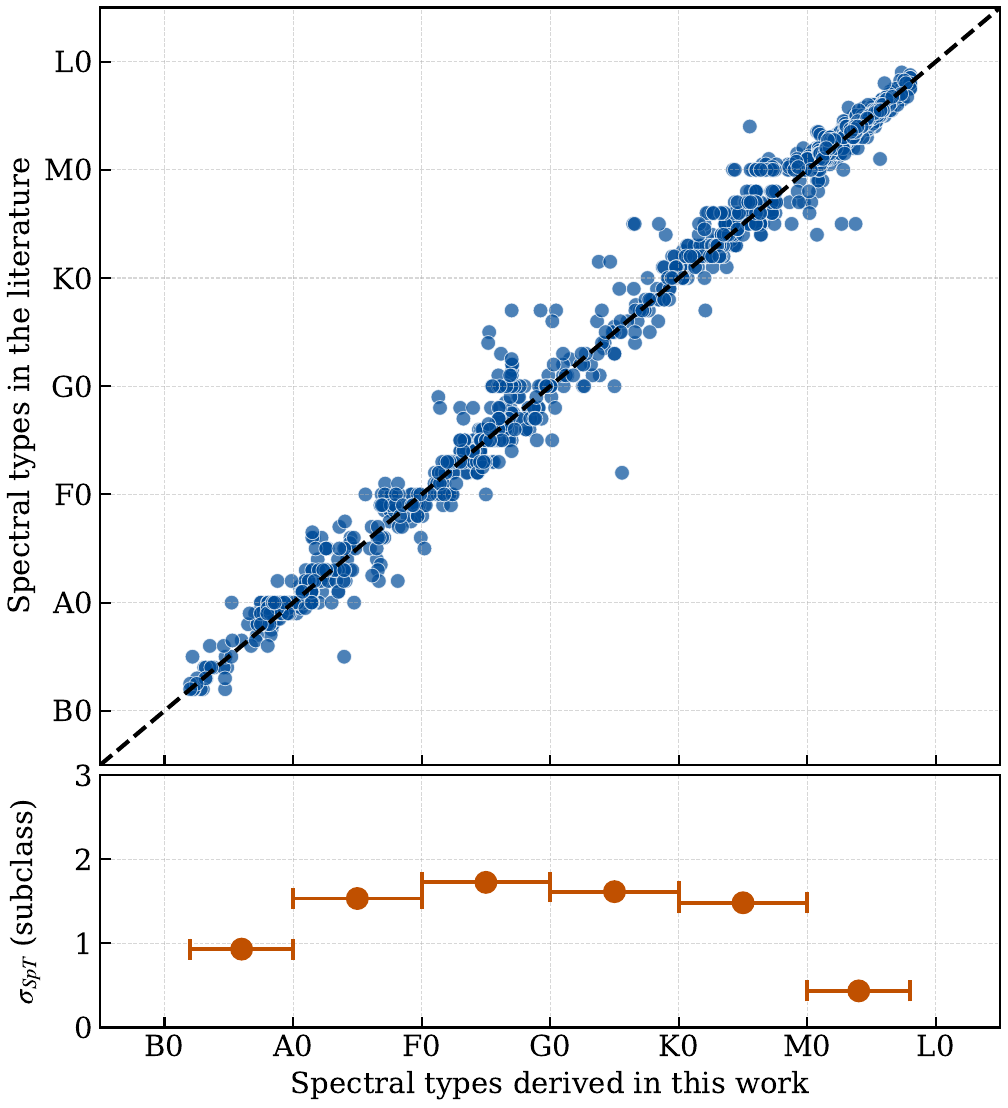}
\caption{Comparison of spectral types (SPT) derived in this study with literature values for 2,155 common objects (upper panel). The lower panel displays the robust standard deviation of SPT differences  across spectral types, with error bars indicating the range of SPT bins employed to compute these dispersions.}\label{Fig:upscodif}
\end{center}
\end{figure}

 \begin{itemize}
\item {\bf Step 1}: We interpolate the spectral templates onto a grid with a spacing of 0.5 spectral subclasses and use these interpolated templates for fitting. Initially, we redden the templates by applying a visual extinction $A_{\rm V}$, and then normalize the reddened templates using the median flux within the wavelength range of 7420\,\AA\ to 7520\,\AA. The target spectra are normalized in the same manner. The best-fit template is determined by minimizing the reduced chi-squared ($\chi^{2}_{r}$). Ultimately, this process yields the optimal values for the two parameters: SPT and $A_{\rm V}$.

\item {\bf Step 2}: In this step, we use the best-fit values for SPT and $A_{\rm V}$ obtained in Step~1 as initial input parameters. The target 7,923 spectra are refitted, and  $\chi^{2}_{r}$ is minimized through an automated iterative procedure. This process iteratively refines the values for SPT and $A_{\rm V}$ to find their optimal values. During each iteration, the templates are interpolated as needed to ensure the best possible fit.
 \end{itemize}

Figure~\ref{Fig:fitexample} shows examples of our spectral fitting for 11 candidate members with spectral types ranging from approximately B5 to M5. Out of the 7,923 sources with Gaia XP spectra, 7,847 can be classified.  The remaining 76 spectra are either featureless or exhibit ripple-like patterns or artifacts and therefore cannot be classified.

Given the low spectral resolution of Gaia XP, the procedure for spectral classification used in this work does not account for the effect of veiling in many accreting young stars. Veiling introduces additional continuum emission, which  dilutes photospheric absorption lines typically used for spectral classification and complicates the determination of intrinsic stellar properties, such as effective temperature   \citep{1998ApJ...509..802C,2004ApJ...616..998W, 2014ApJ...786...97H, 2021ApJ...908...49F, 2020ApJ...904..146F}. Specifically, for K/M type young stars, neglecting veiling effects may result in systematically assigning earlier spectral types than are accurate \citep{2020ApJ...904..146F}.  The effect of veiling predominantly impacts younger regions with higher disk fractions. 

{\newrev Among the sources with Gaia XP spectra, 768 exhibit infrared excess. Of these, 96\% (737/768) display detectable H$\alpha$ pseudo-equivalent widths ($PEW_{\rm{H}\alpha}$) as reported in \cite{2023A&A...674A..28F}, while 68\% (523/768) show H$\alpha$ emission lines, marking them as potential accretors. For these H$\alpha$ emitters, their $PEW_{\rm{H}\alpha}$ values are converted to equivalent widths ($EW_{\rm{H}\alpha}$) using the $PEW_{\rm{H}\alpha}–EW_{\rm{H}\alpha}$
  relationship derived by \cite{2025A&A...699A.145D}. Sources with detectable H$\alpha$ emission lines are classified as accreting or non-accreting using the spectral-type-dependent $EW_{\rm{H}\alpha}$
  criteria outlined in \cite{2009A&A...504..461F}. In total, 230 stars are identified as accreting, with their spectral classifications potentially influenced by veiling.}

\begin{figure*}
\begin{center}
\includegraphics[width=2\columnwidth]{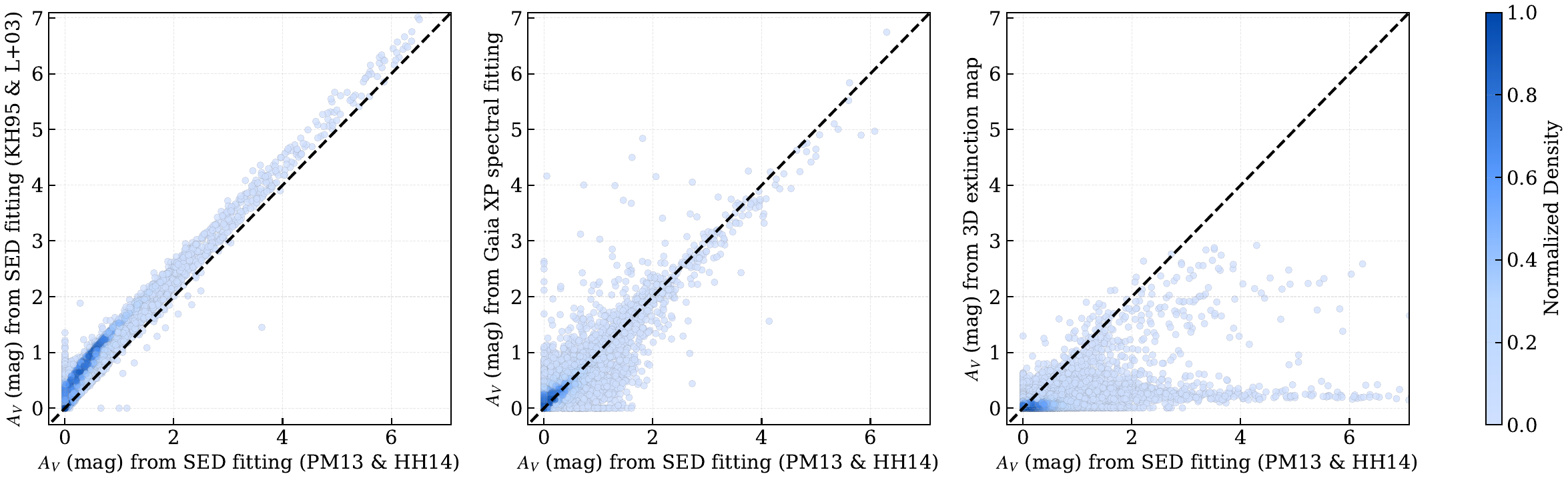}
\caption{Comparison of $A_{\rm V}$ derived from the SED fitting using the two SPT-$T_{\rm eff}$ conversions (left), from the SED fitting using PM13 \&HH14 conversion and the spectral fitting (middle),  from the SED fitting using PM13 \&HH14 conversion and
qthe integrated extinction map \citep{2024A&A...685A..82E}. In each panel, the color is represented by the normalized density.}\label{Fig:Avcom}
\end{center}
\end{figure*}

\begin{figure}
\begin{center}
\includegraphics[width=1.0\columnwidth]{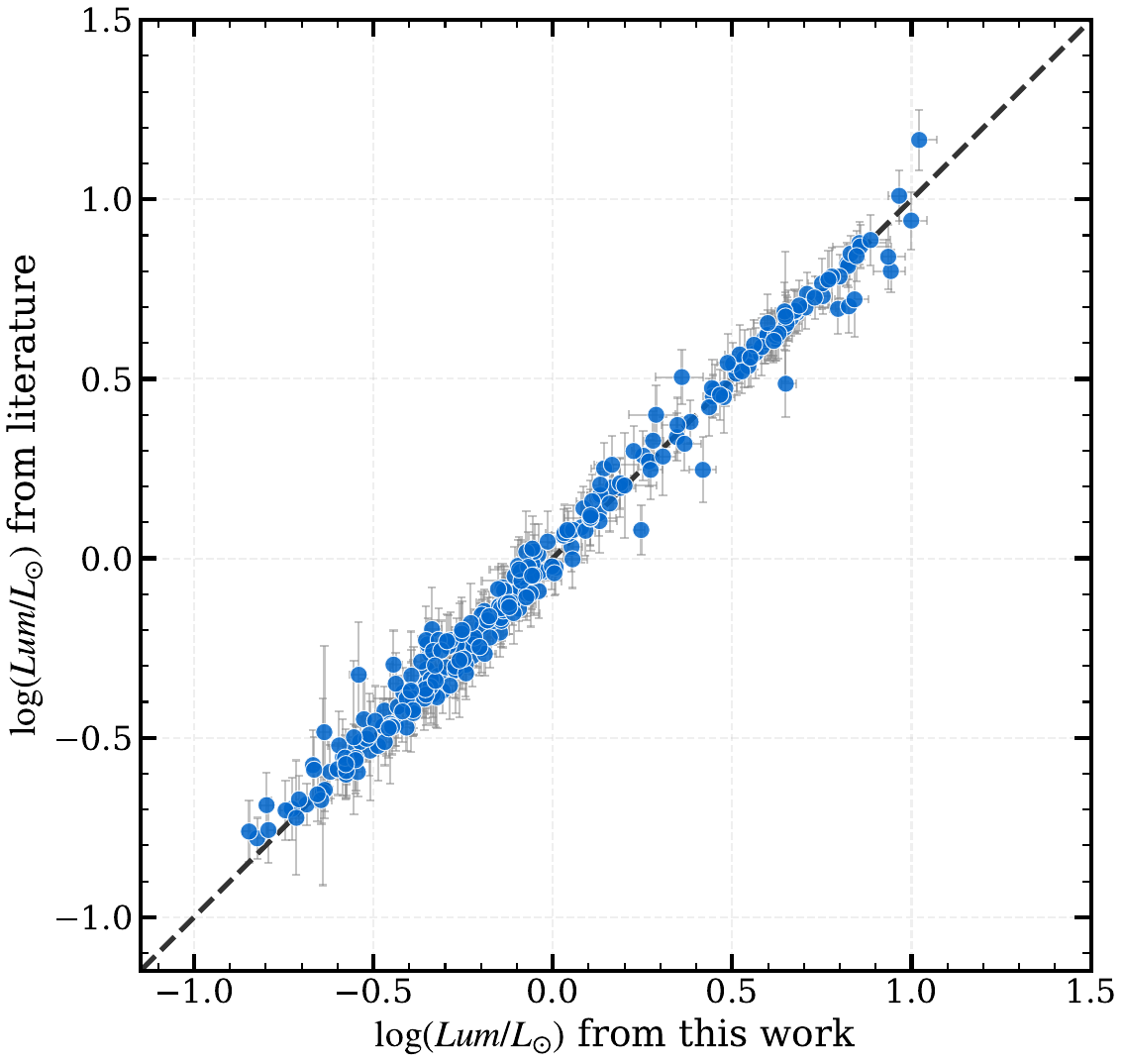}
\caption{Comparison of stellar luminosities as derived in this study and as reported in the literature. The dashed line represents the one-to-one relationship.
}\label{Fig:Lumcom}
\end{center}
\end{figure}

\begin{figure*}
\begin{center}
\includegraphics[width=2.0\columnwidth]{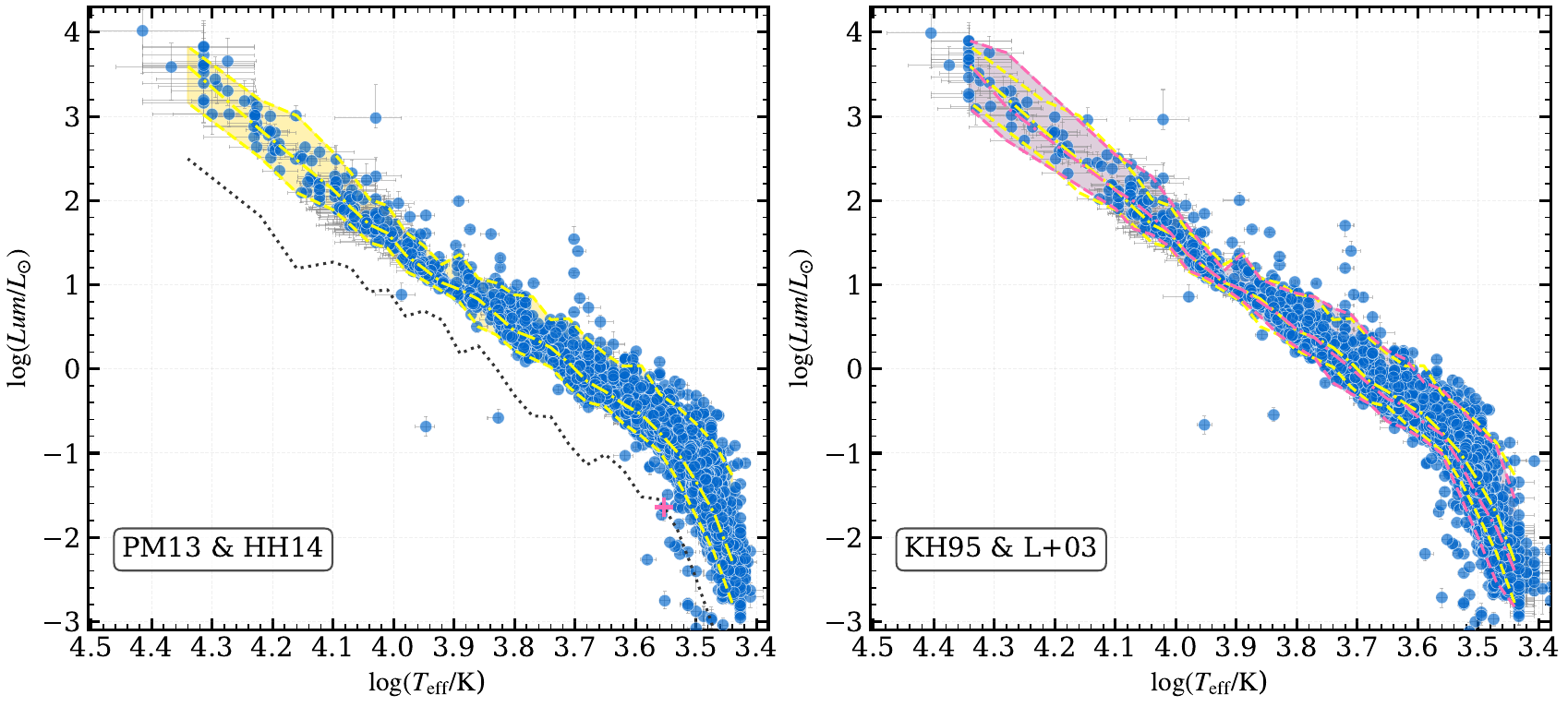}
\caption{H-R Diagram for the candidate members (blue filled circles with errbars) in Sco-cen complex using conversions from PM13 \& HH14 (left) and KH95 \& L+03 (right). In each panel, the median empirical locus for the population is indicated by a dash-dotted line (yellow for the left panel and pink for the right panel), with the 95\% confidence interval (yellow dashed lines for the left panel, and pink dashed lines for the left panel)  for the median value shown as a shaded region (golden for the left panel and plum for the right panel) enclosed by dash lines of the same color. For comparison, the median empirical locus (yellow dash-dotted line) and its 95\% confidence interval (yellow dashed lines) from the left panel are also displayed in the right panel. In the left panel, the black dotted line show the 5$\sigma$ deviation from the median empirical locus. There are 16 sources below the 5$\sigma$ deviation, 15 ones are harboring disks and the other one without a disk could be a field contaminator (marked as the pink plus symbol). 
}\label{Fig:Two_Teff}
\end{center}
\end{figure*}

\subsection{Extinction and Stellar Luminosity}\label{sect:SED}
 We perform a broader spectral energy distribution (SED) analysis to simultaneously determine the extinction and stellar luminosity ($Lum$) for each young star in our sample. The SED for each source is constructed using the optical and infrared photometry described in Section \ref{sec:photometry}. 
 For each source, we calculate its distance and associated standard deviation by inverting 10$^{6}$ 
  samples drawn from its parallax normal distribution.

Our SED fitting process aligns the optical and near-infrared measurements with reddened model atmospheres, using effective temperatures ($T_{\rm eff}$) converted from the spectral classification.
{\newrev The extinction law follows the formulation in \citet{1989ApJ...345..245C}, with a total-to-selective extinction value of $R_{\rm V} = 3.1$. }
To test how the $SpT-T_{\rm eff}$ conversion affects the derived parameters, we use two $SpT-T_{\rm eff}$ conversion: the relation  provided by \citealt{2017AJ....153..188F}, hereafter PM13\&HH14 conversion, which use conversions from \citet{2013ApJS..208....9P} for stars earlier than M4 and from \citet{2014ApJ...786...97H} for stars later than M4; and separately the combination of conversions from (KH95) \citet{1995ApJS..101..117K} for stars earlier than M0 and from (L+03) \cite{2003ApJ...593.1093L} for stars later than M0.
  We use the spectral types derived in this work when there are available and reliable. When not available in this work, we adopt spectral types in the literature.
The model atmospheres used are the BT-Settl models \citep{2011ASPC..448...91A} with solar abundances from \citet{2009ARA&A..47..481A}. During the SED fitting, for diskless sources we include photometry in both the optical bands and the infrared $J$, $H$, $K_S$, W1, and W2 bands. For sources with disks, we use photometry in the optical bands and the 2MASS-J band. 
The uncertainty on the $Lum$ for each source is estimated from its uncertainty of the spectral type and distance.

\begin{figure*}
\begin{center}
\includegraphics[width=2\columnwidth]{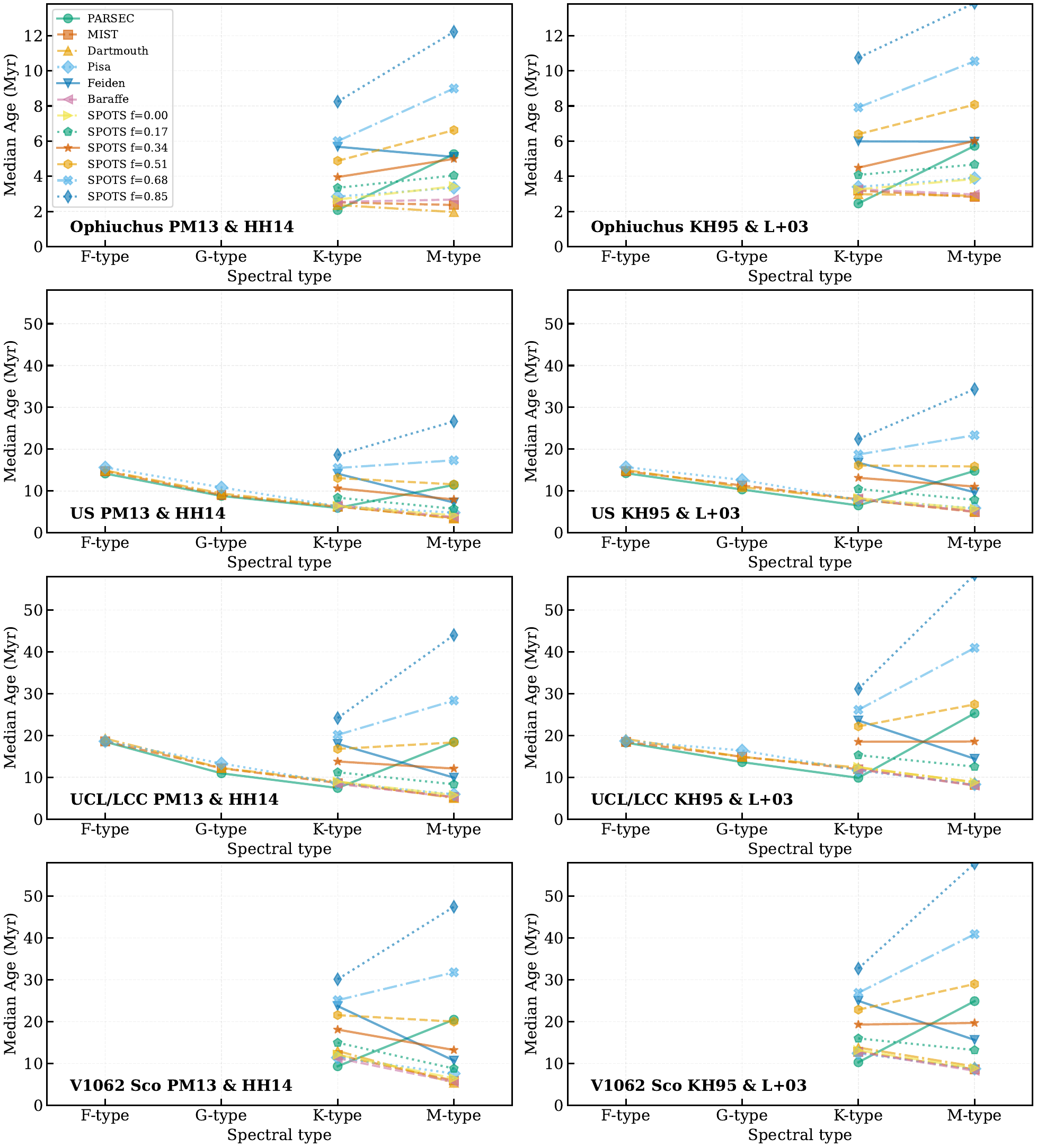}
\caption{Median ages of UCL/LCC (top) and US (bottom)  within 4 spectral ranges: F type but later than F5, G types, K types, M types (M0$-$M5), derived with different stellar evolutionary tracks.       }\label{Fig:median_age}
\end{center}
\end{figure*}

{\newrev
To investigate how varying $R_{\rm V}$  affects extinction and stellar luminosity results, we additionally performed SED fitting with $R_{\rm V}=4.0$. A change in $R_{\rm V}$ from 3.1 to 4.0 leads to lower $A_{V}$ values and higher $Lum$  values. For $A_{V}$=$1$, $3$, and $5$ at $R_{\rm V}=3.1$, the median differences in $A_V$ would be 0.2, 0.8, and 1.3 mag, leading to median differences in $Lum$ of  0.001, 0.04, and 0.07 dex. Notably, the majority (83\%) of our sources  exhibit $A_{V}<=1$, where the luminosity difference is negligible. For sources in the Ophiuchus region, where objects may be more deeply embedded in the molecular cloud, the median $A_{V}$ at  $R_{\rm V}=3.1$ is approximately 2 mag, with only 9\% of these sources showing $A_{V}>$5~mag. Collectively, these findings suggest our sources are likely to be minimally affected by $R_{\rm V}$ variations.}

\section{Results} 
\subsection{Spectroscopic Sample}
We determined the spectral type of the candidate members of  with the Gaia XP spectra as described in \sect~\ref{sect:class}. Of the sources with Gaia XP spectra classified in this study, 2,155 sources have spectral-type estimates in the literature \citep{2012ApJ...746..154P,2016MNRAS.461..794P,2021ApJ...908...49F,2022AJ....163...64E,2022AJ....163...26L}. Figure~\ref{Fig:upscodif} (upper panel) shows the comparison between the spectral types determined in this study and those reported in the literature. Approximately 79\% of the sample agrees within one spectral subclass and 91\% agree within two subclasses. For sources with spectral types later than M0, which account for 83\% of the total samples with spectral types, these percentages increase to 94\% and 99\%, respectively. To quantify the standard deviations of spectral type (SpT) differences, we divided the sample into 7 SpT bins, each spanning one spectral class. 
For each bin, we computed the standard deviation of the SpT differences between this study and literature values. The results are shown in the lower panel of Figure~\ref{Fig:upscodif}. The smallest standard deviation (0.4 subclasses) occurs for M-type sources, while for  other spectral types, the standard deviations range between 0.9 and 1.7 subclasses. This  is generally consistent with the uncertainty for our spectral templates (see Appendix~\ref{Appen:template}), 
taking that the uncertainty on spectral types in the literature is comparable to ours.

In total, 84\% of previously identified members (8,861 sources) have spectral types, of which 64\% (5,692 of 8,861) are spectroscopically classified here. 
In Table~\ref{tabe_UpperSco}, we list the 8,861 sources 
with the spectral types from this work or the literature. 
Among the sources with the spectral types, 8,846 have photometry in more than three bands which can be used for the SED fitting (see \sect~\ref{sect:SED}). 
Figure~\ref{Fig:spsample} shows the numbers and the proportion of these sources across individual Gaia G magnitude interval. For the sources with Gaia G-band magnitude brighter than 17~mag, more than 92\% of the candidate members have been observed with spectroscopy.
In Table~\ref{tabe_UpperSco}, we list the distance, $T_{\rm eff}$, log~$Lum$, and $A_{\rm V}$ of these sources from the PM13 \&HH14  $SpT-T_{\rm eff}$ conversions.

The left panel of Figure \ref{Fig:Avcom} compares the $A_{\rm V}$ values derived from SED fitting using two different SpT-$T_{\rm eff}$ conversions. The $A_{\rm V}$ values obtained using the KH95 \&L+03 conversion are systematically higher than those derived from the PM13 \& HH14 conversion. The discrepancy arises because, for the same spectral type, temperatures derived from the KH95  \& L+03 conversion are systematically 1\%–6\% higher than those from the PM13 \& HH14 conversion across spectral types G3 to M8.
The middle panel of Figure \ref{Fig:Avcom} compares the $A_{\rm V}$  values obtained from SED fitting using the PM13 \& HH14 conversion with those derived from Gaia XP spectral fitting. Generally, the $A_{\rm V}$ values from both methods are in very good agreement, showing no systematic bias. 

The right panel of Figure \ref{Fig:Avcom} compares the $A_{\rm V}$ values obtained from SED fitting using the PM13 \& HH14 conversion with those derived from the 3D extinction map with an angular resolution of 14$^\prime$  presented by \cite{2024A&A...685A..82E} and constructed using the extinction values from \cite{2023MNRAS.524.1855Z}. For each star, we retrieved the integrated extinction up to its distance.
Overall, the $A_{\rm V}$ values from the extinction map are systematically lower than those derived from the SED fitting, suggesting that these stars are primarily reddened by the local interstellar medium (ISM) rather than by foreground dust. This discrepancy indicates a caution when dereddening young stars using extinction maps, as it may underestimate the total extinction affecting these objects.

Figure~\ref{Fig:Lumcom} compare the stellar luminosity obtained in this work with the PM13 \& HH14 conversion  and  in the literature \citep{2013ApJS..208....9P,2016MNRAS.461..794P}. Both the stellar luminosities are derived using the same  $SpT-T_{\rm eff}$ conversion. The $Lum$ from the literature have been scaled to the distances of individual sources. The $Lum$ in Figure~\ref{Fig:Lumcom} from two resources 
agree with each other with the  mean difference of 0.029~dex. We also compare the $Lum$ derived using the KH95 \& L+03 conversion with the literature values, and the difference on the $Lum$ is 0.040~dex. The relatively higher difference is due to that the $Lum$ is derived using different $SpT-T_{\rm eff}$ conversions.

\subsection{Hertzsprung-Russell Diagram}\label{Sect:HRD}
With the derived \Teff\ and bolometric luminosities, we place the candidate members of the Sco-Cen complex in the H-R diagram in Figure~\ref{Fig:Two_Teff}. The two H-R diagrams in the left and right panels are derived using the PM13 \& HH14  and  KH95 \& L+03 conversions, respectively. While the median empirical locus from the two conversions are generally consistent with each other when \Teff\ is higher than $\sim$6300~K, they systematically deviate from each other at the lower temperature, resulting in a systematic age discrepancy between them.

Figure~\ref{Fig:Two_Teff} shows that 16 sources have $Lum$ lower than median empirical locus by more than 5$\sigma$. Of these, 15 show strong infrared excesses and thus have circumstellar disks. The presence of circumstellar disks may lead to variable or permanent obscuration of the star and variable accretion, both of which can affect the estimate of stellar luminosity.
These sources are considered members of the Sco-Cen complex.  In addition, some edge-on disks are likely either too faint for Gaia or have large RUWE or parallax uncertainties and would be excluded from this study. The remaining outlier (ID\# 8562 in Table~\ref{tabe_UpperSco}) does not show any infrared excess and could be contamination by a field star.

\begin{figure*}
\begin{center}
\includegraphics[width=2\columnwidth]{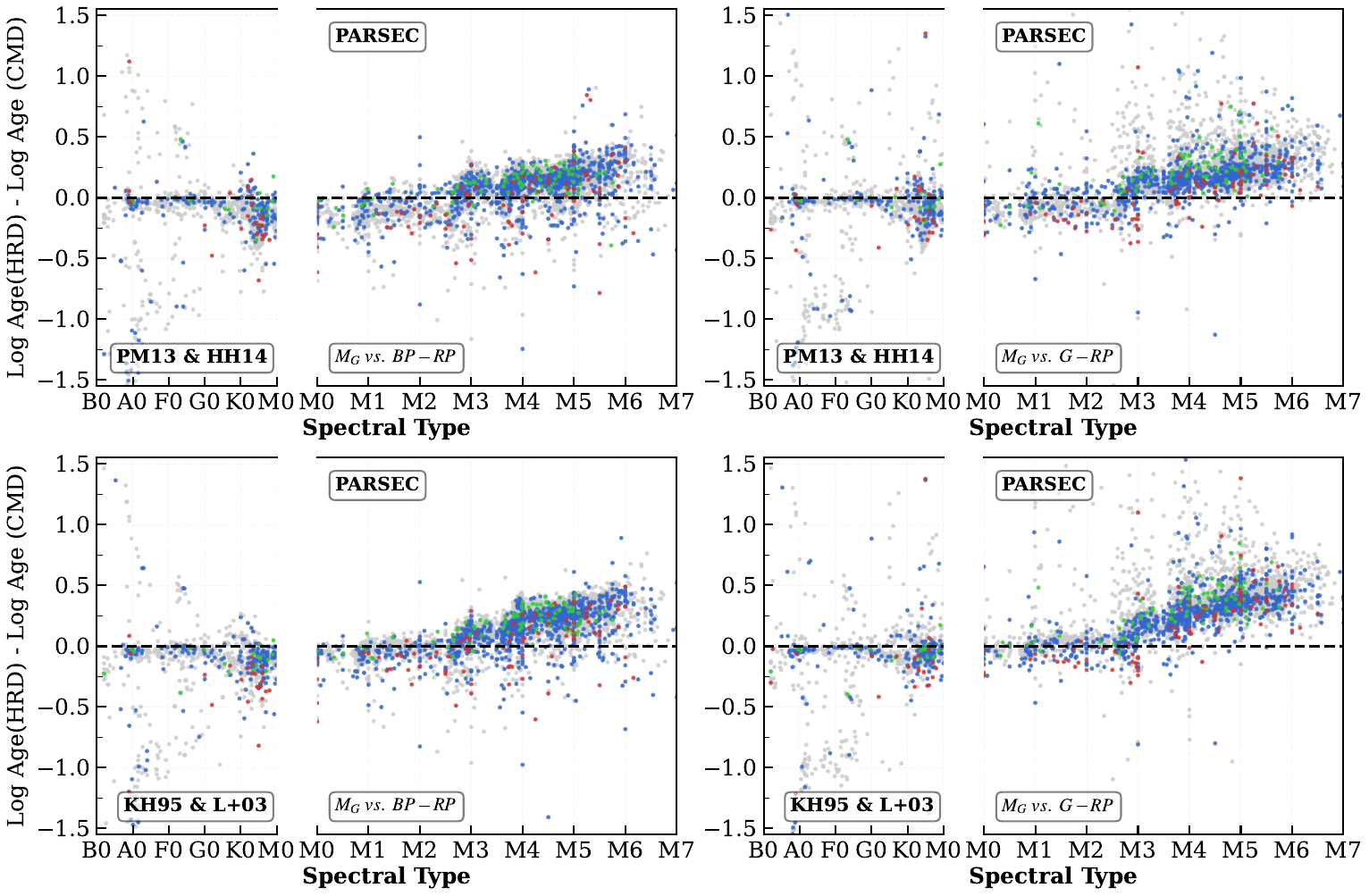}
\caption{The age differences between the HRD-based and CMD-based estimates for the Ophiuchus (red dots), US (blue dots), UCL/LCC (gray dots), and V1062 Sco (green dots) groups are plotted as a function of spectral type. Ages are derived using the PARSEC models in conjunction with two sets of spectral type-to-effective temperature conversions: PM13 \& HH14 (top panels) and KH95 \& L+03 (bottom panels).
    }\label{Fig:agedif_Parsec}
\end{center}
\end{figure*}

\begin{figure*}
\begin{center}
\includegraphics[width=2\columnwidth]{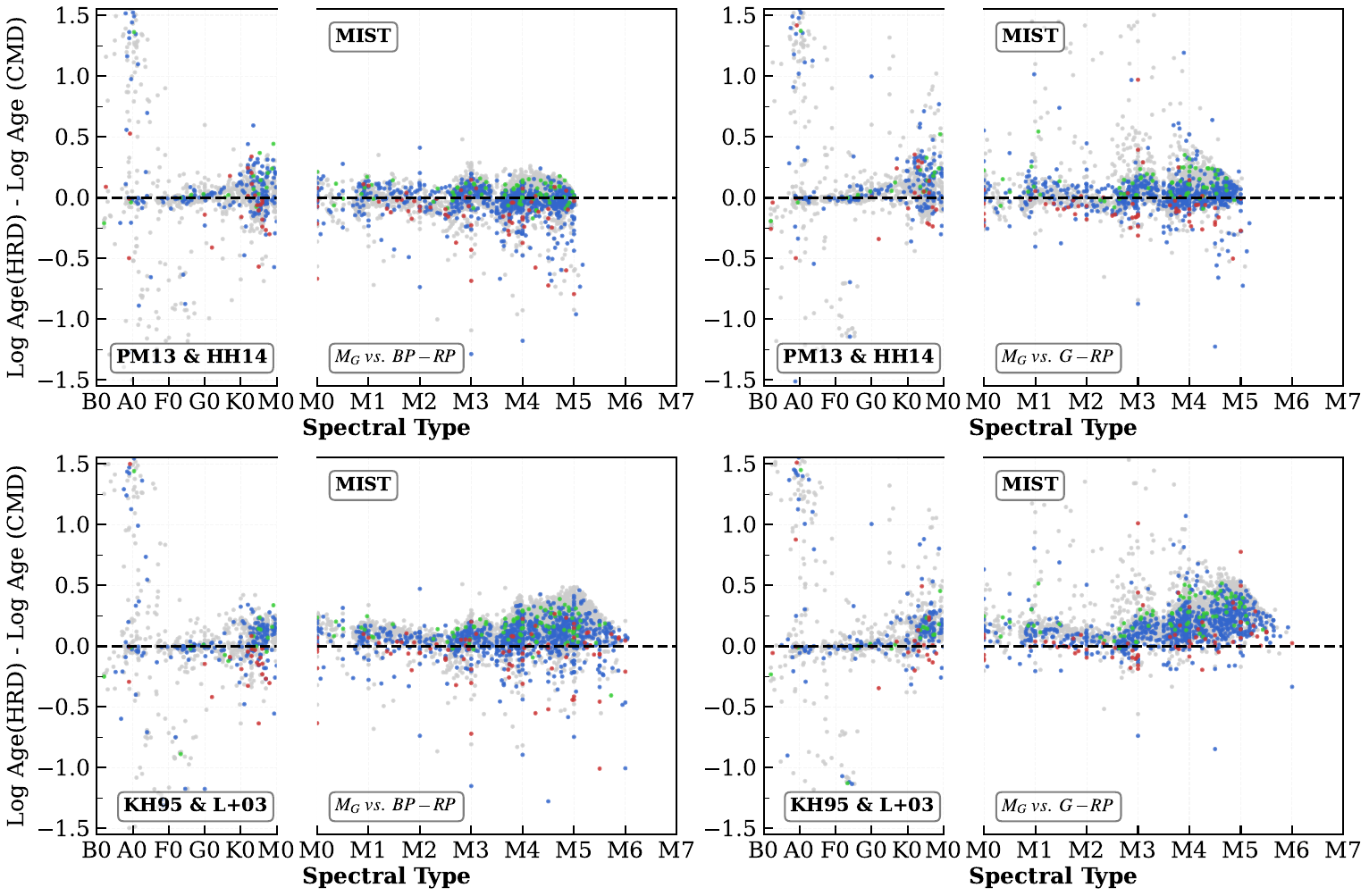}
\caption{Same as Fig.~\ref{Fig:agedif_Parsec} but using the  MIST models.}\label{Fig:agedif_MIST}
\end{center}
\end{figure*}

\begin{figure*}
\begin{center}
\includegraphics[width=2\columnwidth]{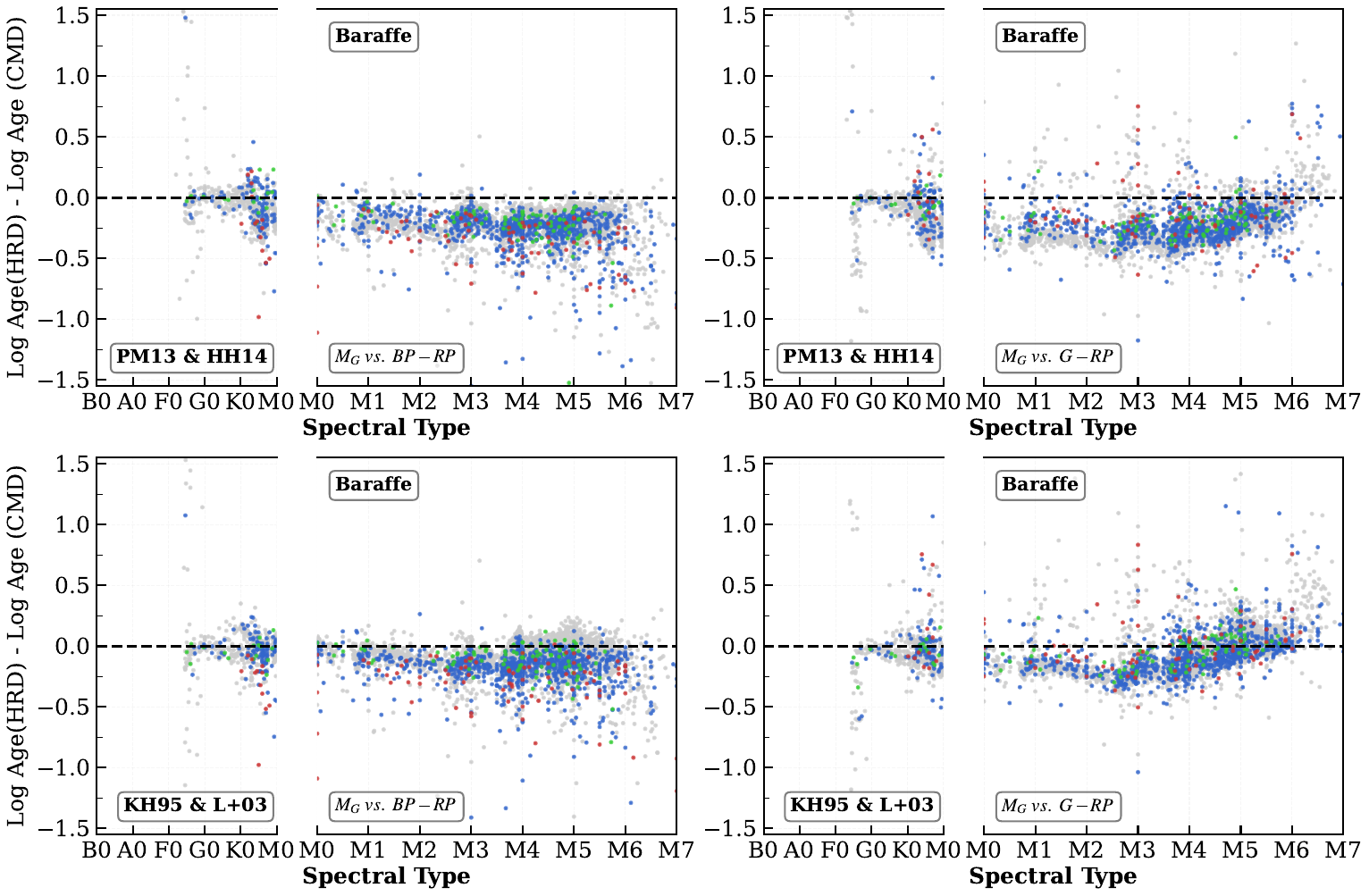}
\caption{Same as Fig.~\ref{Fig:agedif_Parsec} but using the Baraffe models.
}\label{Fig:agedif_Baraffe}
\end{center}
\end{figure*}

\begin{figure*}
\begin{center}
\includegraphics[width=1\columnwidth]{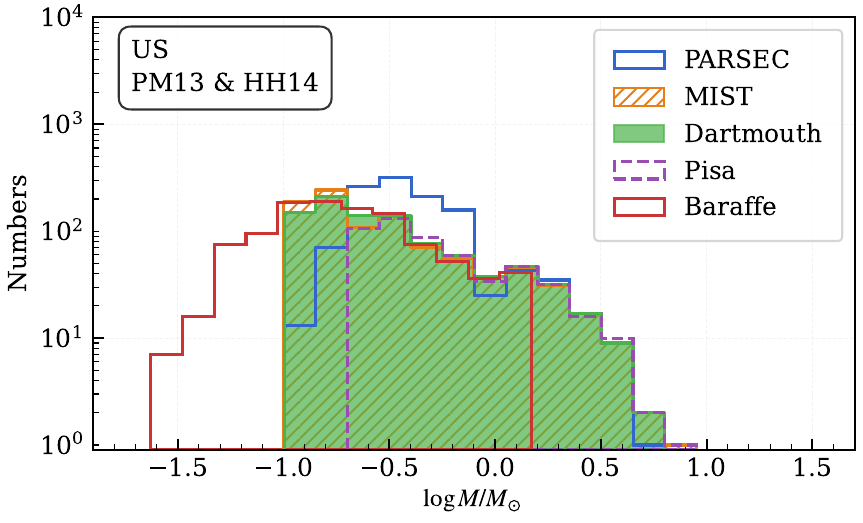}
\includegraphics[width=1\columnwidth]{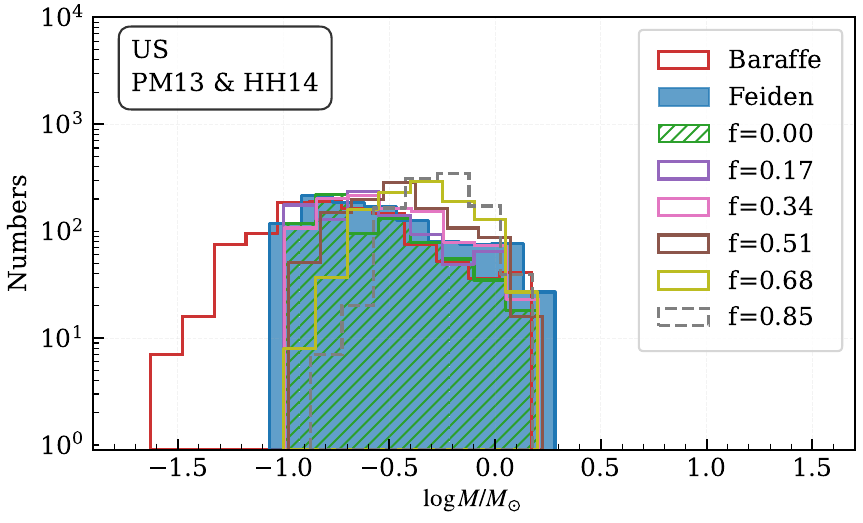}
\includegraphics[width=1\columnwidth]{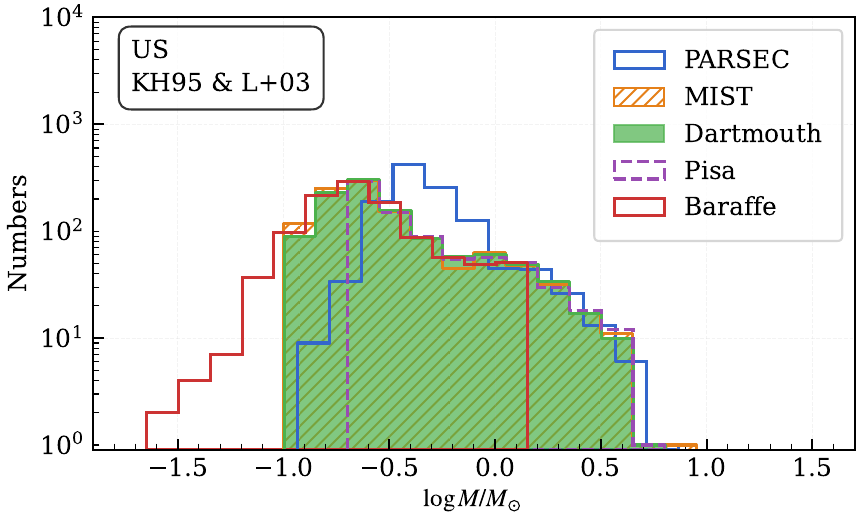}
\includegraphics[width=1\columnwidth]{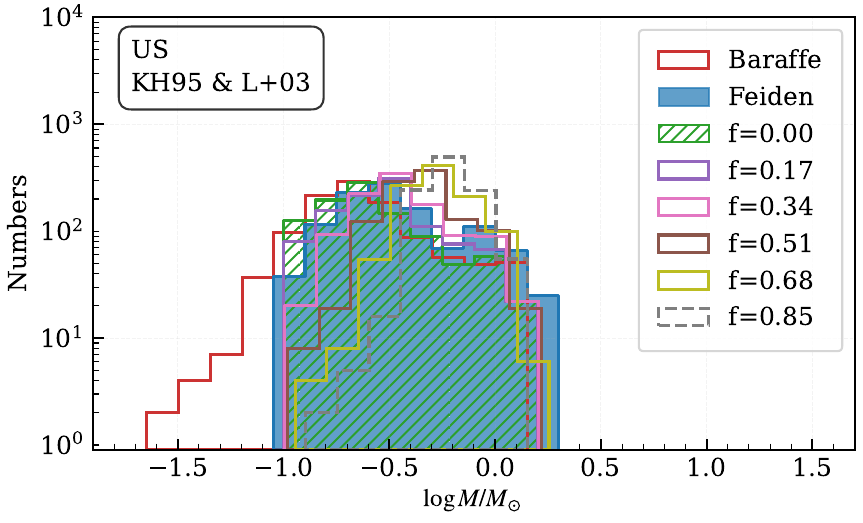}
\caption{Mass distributions of the sources in the US  group derived with different evolutionary models. 
}\label{Fig:mass_us}
\end{center}
\end{figure*}

\begin{figure*}
\begin{center}
\includegraphics[width=1\columnwidth]{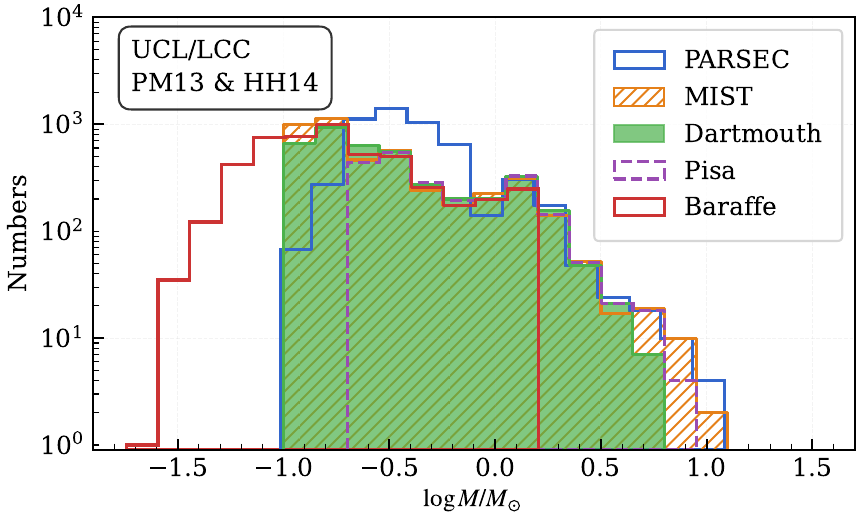}
\includegraphics[width=1\columnwidth]{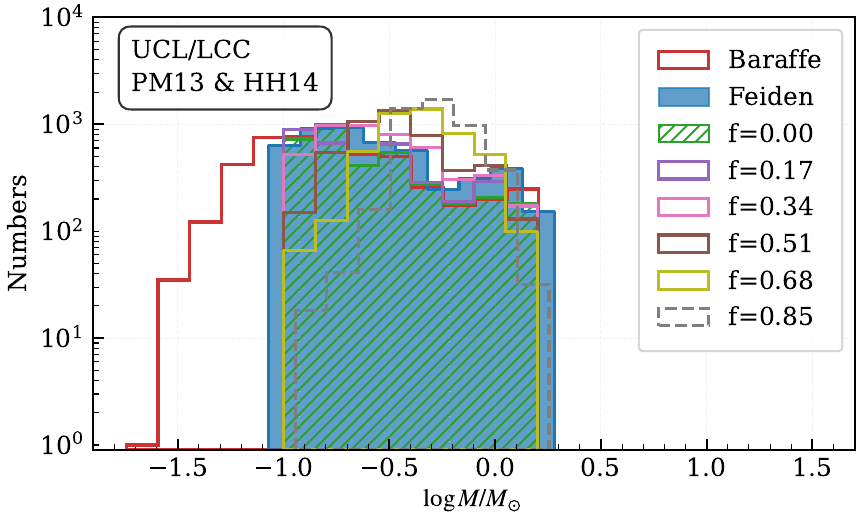}
\includegraphics[width=1\columnwidth]{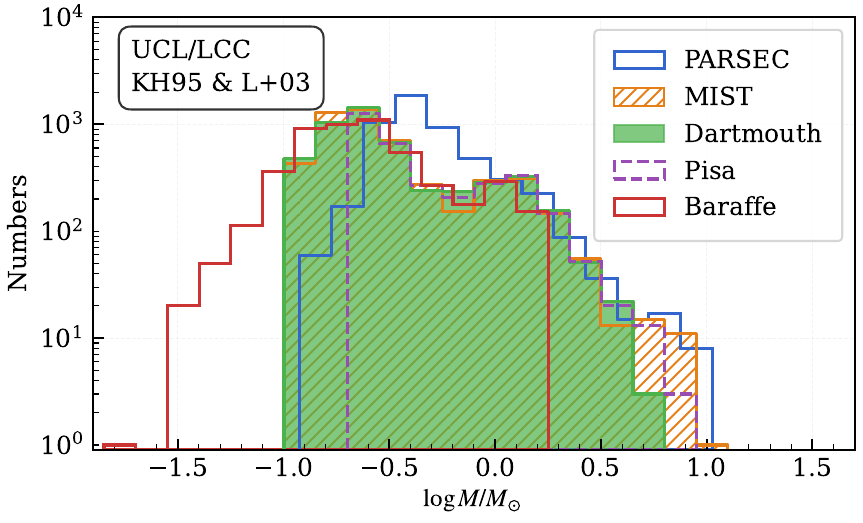}
\includegraphics[width=1\columnwidth]{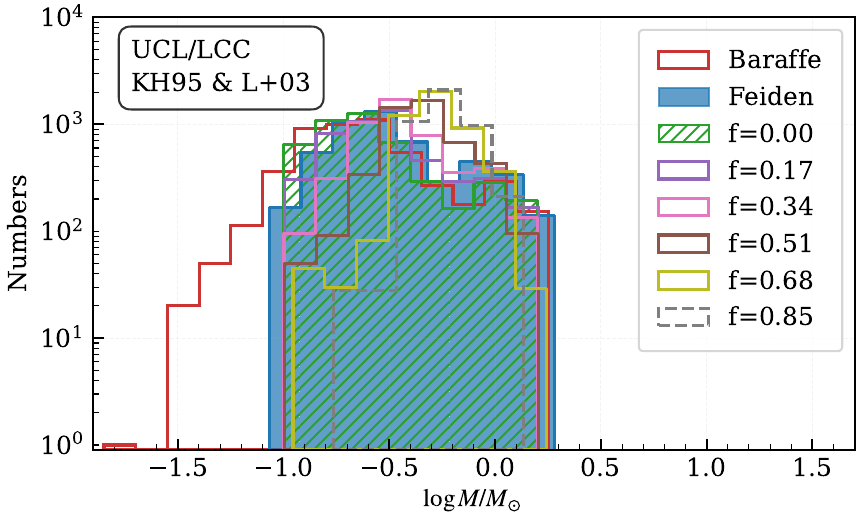}
\caption{Same as Figure~\ref{Fig:mass_us} but for the UCL/LCC group.
}\label{Fig:mass_ucl}
\end{center}
\end{figure*}

\setcounter{table}{1}
\begin{table*}
\scriptsize
\renewcommand{\tabcolsep}{0.04cm}
{\newrev
\caption{\newrev Parameters for evolutionary models }\label{Table:ev_models}
\begin{center}
\begin{tabular}{lcccccc}
\hline
\hline
Models & Mass Range & atmosphere models   \\
       & [$M_{\odot}$]  & \\
\hline 
PARSEC 1.2S&0.1-350  &ATLAS9$^{\alpha}$ models for $T_{\rm eff}>6500$~K and BT-Settl$^{\beta}$  models for $T_{\rm eff}<5500$~K    \\
           &    &Interpolation of ATLAS9 and BT-Settl models for 6500~K$>T_{\rm eff}>5500$~K   \\
MIST      &0.1-300   & PHOENIX$^{\gamma}$ models for 10,000~K$>T_{\rm eff}>3000$~K   &  \\
           &   &Using ATLAS9$^{\alpha}$ models (50,000~K$>T_{\rm eff}>3500$~K) when PHOENIX ones are unavailable   &  \\
Dartmouth &0.1-4     &PHOENIX$^{\gamma}$ models for 10,000~K$>T_{\rm eff}>2000$~K   &  \\
          &          &ATLAS9$^{\alpha}$  models for $>T_{\rm eff}>10,000$~K   &  \\
Pisa      &0.2-7.0  &PHOENIX$^{\theta}$ models for 10,000~K$>T_{\rm eff}>3000$~K    &  \\
          &         &Using ATLAS9$^{\alpha}$ models (50,000~K$>T_{\rm eff}>10,000$~K)  when PHOENIX ones are unavailable     &  \\
Baraffe   &0.01-1.4  & BT-Settl CIFIST2011\_2015$^{\delta}$ models &  \\
Feiden Magnetic model &0.085-1.7 &PHOENIX$^{\gamma}$ models     &  \\
SPOTSs    &0.1-1.3  &NextGen$^{\mu}$ models for  M$<0.4~M_{\odot}$   &  \\
          &        &ATLAS9$^{\alpha}$ models for  M$>0.6~M_{\odot}$  &  \\
          &        &Interpolation of NextGen and ATLAS9 ones for $0.6~M_{\odot}>$M$>0.4~M_{\odot}$  &  \\          
          \hline
\end{tabular}
\end{center}
\tablecomments{$\alpha$: \citet{2003IAUS..210P.A20C}; $\beta$:  \citet{2012EAS....57....3A}; $\gamma$: \citet{1999ApJ...512..377H} and \citet{1999ApJ...525..871H}; $\theta$: \citet{2005ESASP.576..565B}; $\delta$: \citet{2015A&A...577A..42B} and \citet{2016sf2a.conf..223A}; $\mu$: \citet{1997ARA&A..35..137A}
}
}
\end{table*}

\subsection{Comparing ages from different models}\label{Sect:ISO}

To estimate the ages of the populations in the Sco-Cen complex, we compare these sources in the H-R diagrams with model isochrones from 
the PARSEC models \citep{2012MNRAS.427..127B}, the MIST models \citep{2016ApJ...823..102C}, the Dartmouth models \citep{2008ApJS..178...89D}, the Pisa models \citep{2011A&A...533A.109T}, the Baraffe models \citep{2015A&A...577A..42B}, the Feiden magnetic models \cite{2016A&A...593A..99F}, and 
the  SPOTS models with six 
spot surface covering fraction $f_{\rm spot}=$0, 0.17, 0.34, 0.51, 0.68, and 0.85 \citep{2020ApJ...891...29S}. 
{\newrev The individual models, including their mass ranges and atmosphere models, are summarized in Table~\ref{Table:ev_models}.
 For each model, we adopt the solar-metallicity evolutionary tracks and  linearly interpolate the original isochrones onto a logarithmic age grid with a spacing of 0.001~dex to construct densely sampled isochrone sequences.  For each star, its optimal age and mass are then derived by matching to the closest isochrone within these densely spaced grids in the H-R diagram.}

\begin{figure*}
\begin{center}
\includegraphics[width=2\columnwidth]{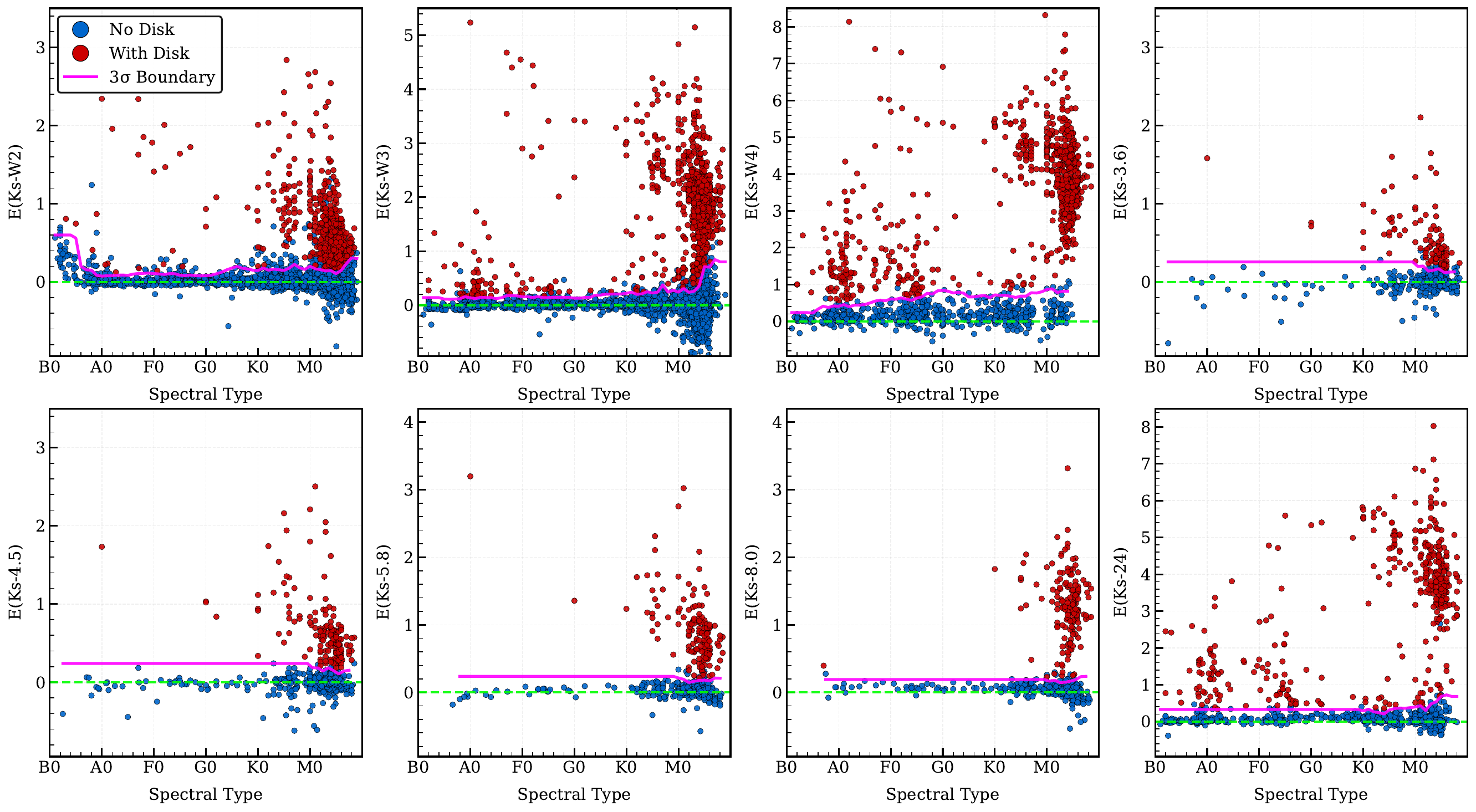}
\caption{Extinction-corrected infrared color excess vs. Spectral type for the sources in the Sco-Cen complex. In each panel, the blue filled  circles are for the sources without the color excess and the red filled circles are for the ones with the color excess, and the magenta solid line marks the 3$\sigma_{Excess}$ level. }\label{Fig:EKIR}
\end{center}
\end{figure*}

Figure~\ref{Fig:median_age} compares the median ages for 4 groups (Ophiuchus, US, UCL/LCC, and the V1062~Sco Group)\footnote{The group classification is from \cite{2022AJ....163...24L}} derived with different models with the two SPT-$T_{\rm eff}$ conversions. The sources are divided into 4 SPT bins: F types but later than F5, G types, K types, and M0-M5 types. For the Baraffe, Feiden, and SPOTS models, the median ages are only estimated for the K and M type bins due to the mass range of these models. For Ophiuchus, the youngest of the four groups,  the median ages are only estimated for the K and M type bins for all the models due to the small number of F and G members.

For the three older groups (US, UCL/LCC, and V1062~Sco group), the median ages calculated using the MIST, Dartmouth, PISA, and Baraffe models {\newrev exhibit a trend of decreasing age to later spectral types.}
Within F-type, G-type, and K-type bins, the median ages derived with the PARSEC models are comparable with the ones from the above models. However, unlike the other models, the median age of PARSEC models within the M-type bin are much older than the median age of the K-type bin.  For the Ophiuchus group, the median ages within the K-type and M-type bins are comparable for the MIST, Dartmouth, PISA, and Baraffe models. 

The Feiden models consistently yield ages that are approximately 1.5 to 2 times older for the four groups within K-type and M-type bins, compared to the ages derived from the MIST, Dartmouth, PISA, and Baraffe models \citep[consistent with results found by][]{2016A&A...593A..99F}. For the SPOTS models, the median ages within both K-type and M-type bins increase with greater spot coverage. Notably, for lower spot coverages, the median age in the M-type bin is smaller than that in the K-type bin. However, as spot coverage increases, the median age in the M-type bin becomes older than that in the K-type bin.

{\newrev 
The stellar ages derived here do not account for the effects of metallicity, binarity, or uncertainties in SpT and $Lum$. An increase in the metallicity of the adopted evolutionary models would lead to older age estimates, while binarity could result in slightly younger ages. Including the uncertainties in SpT and $Lum$ has negligible impact on the derived ages for individual stars. A detailed discussion of these effects is provided in Appendix~\ref{Appen:age_effect}.
}

\subsection{Comparing Ages derived with HRDs and CMDs}\label{Sect:ISO}

The PARSEC, MIST, and Baraffe models have published their stellar evolutionary tracks within the Gaia photometric system, which are frequently employed to gauge the isochronal ages of stellar populations in the Gaia color-magnitude diagram (CMD), in particular the PARSEC models \citep[see e.g. ][]{2020ApJ...904..196T,2021ApJ...923...20P,2023A&A...678A..71R,2024NatAs...8..216M}. However, until now, no systematic comparison has been made regarding potential differences in age determinations between Hertzsprung-Russell diagrams (HRDs) and CMDs. In Figures~\ref{Fig:agedif_Parsec}, \ref{Fig:agedif_MIST}, and \ref{Fig:agedif_Baraffe}, we present a comparative analysis of ages derived from HRDs versus those from CMDs ($M_{\rm G}$ vs. BP-RP and $M_{\rm G}$ vs. G-RP) using the three aforementioned models\footnote{\newrev 
Consistent with methods used for H-R diagrams, the original isochrones of individual models in CMDs have been regridded to a finer age spacing of 0.001 dex, producing high-resolution isochrone sets. Subsequently, for each star, its optimal age and mass are identified by aligning with the nearest isochrone within these refined grids in the CMDs.}  
. 

\noindent\textbf{PARSEC Models:}\\
Comparisons of age estimates are conducted for four groups—Ophiuchus, US, UCL/LCC, and V1062 Sco—using two different SPT-$T_{\rm eff}$ conversion methods. When using the PARSEC models, a clear trend emerges in the comparison between HRD-based and CMD-based ages:
\begin{itemize}
    \item For spectral types earlier than K0, HRD-based ages tend to be younger than CMD-based ages.
    \item For spectral types between K0 and M3, the age difference is approximately 0.02–0.04 dex. Notably, within this spectral range, the KH95 \& L+03 conversion results in a smaller age difference compared to the PM13 \& HH14 conversion.
    \item  For spectral types later than M3, HRD-based ages tend to be older than CMD-based ages by between 0.16 and 0.34 dex. Within this range, the KH95 \& L+03 conversion yields a larger age difference compared to the PM13 \& HH14 conversion.
\end{itemize}

\noindent\textbf{MIST Models:}\\
For the MIST models, the age difference between HRD-based and CMD-based estimates is generally smaller compared to those derived from the PARSEC models. Additionally, when using the PM13 \& HH14 conversion, the age differences are smaller than those obtained using the KH95 \& L+03 conversion. 

\noindent\textbf{Baraffe Models:}\\
For the Baraffe models, HRD-based ages tend to be younger than CMD-based ages. Additionally, the age differences are smaller when using the KH95 \& L+03 conversion compared to those obtained using the PM13 \& HH14 conversion.

This analysis highlights the importance of both the chosen stellar evolution model, the stellar atmosphere model, and the SPT-$T_{\rm eff}$ conversion method in determining the consistency of age estimates between HRD and CMD analyses. Each model exhibits distinct trends, underscoring the need for careful consideration of these factors in comparison studies.

\begin{figure*}
\begin{center}
\includegraphics[width=2\columnwidth]{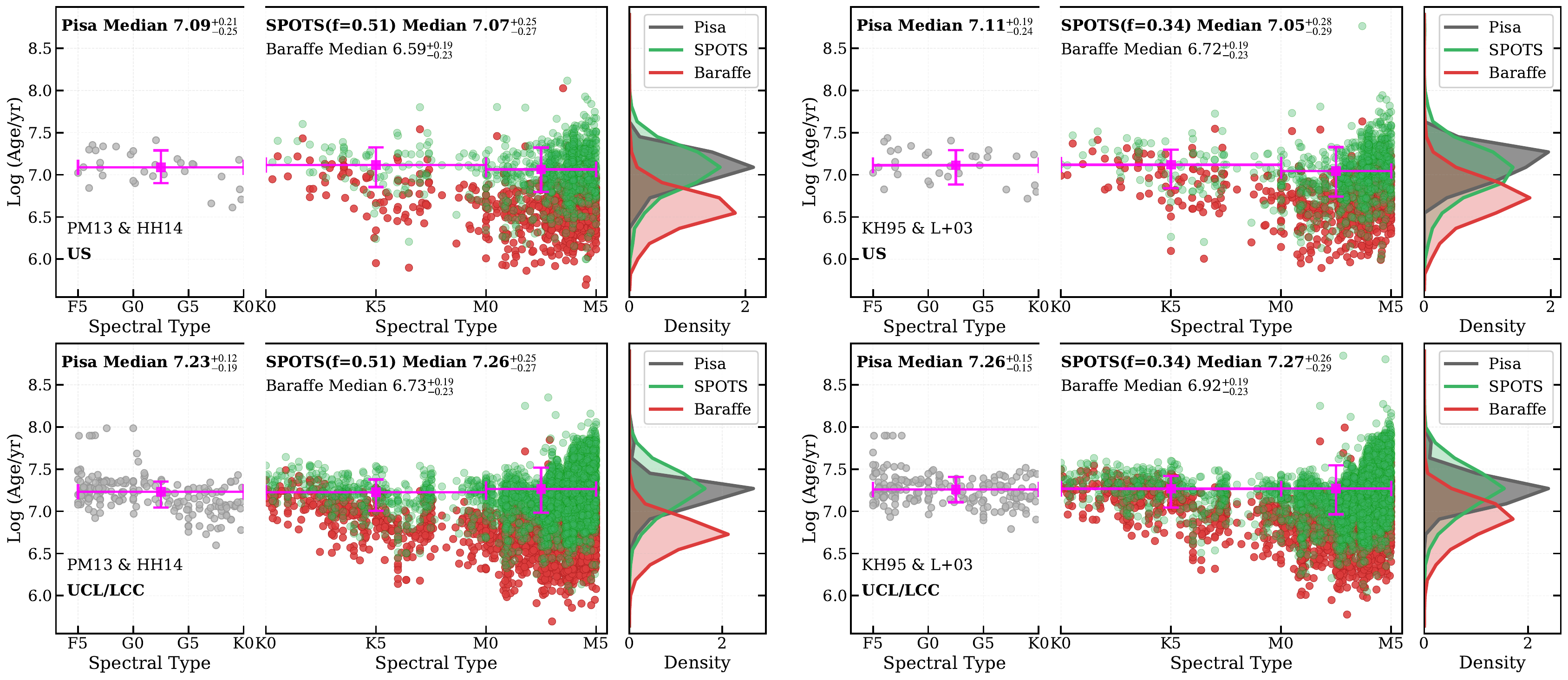}
\caption{{\newrev 
Age estimates for the US (upper panels) and UCL/LCC (lower panels) groups are plotted against spectral type. For F/G-type stars, ages are derived using PISA models (gray); for K/M-type stars, SPOTS (green) and Baraffe (red)  models  are applied. These age calculations incorporate two spectral type-to-effective temperature conversion datasets: PM13 \& HH14 (left panels) and KH95 \& L+03 (right panels). In each panel, the magenta filled squares shows the median age within the F5-G (the Pisa model) and K-M5 type (the SPOTS models) bins, with error bars indicating the 1$\sigma$ uncertainty range. In the right panel, age distributions of the sources derived from the PISA (gray-filled histogram), SPOTS (green filled histogram), and Baraffe (red filled histogram) models are overlaid for comparison.
}
}\label{Fig:agecom_pisa_Spots_US}
\end{center}
\end{figure*}

\subsection{Mass function from different models}\label{Sect:IMF}

In Figures~\ref{Fig:mass_us} and~\ref{Fig:mass_ucl}, we compare the mass functions of the US and UCL/LCC groups using the evolutionary models presented in this work and two SPT-$T_{\rm eff}$ conversion methods. For canonical evolutionary models, the Dartmouth, Pisa, MIST, and Baraffe models yield consistent mass estimates within their respective uncertainties. However, the PARSEC models show agreement with other canonical models only for mass bins above 1\,$M_{\odot}$, while predicting more stars in the 0.4-0.8\,$M_{\odot}$ range and fewer stars below 0.2\,$M_{\odot}$. Uncanonical models, as shown in the right panels of Figures~\ref{Fig:mass_us} and~\ref{Fig:mass_ucl}, generally yield higher masses than their canonical counterparts. Notably, among the SPOTS models, those with greater cool spot coverage systematically produce higher mass estimates compared to models with lower coverage.

\subsection{Disks}
The disk properties of individual sources are initially derived from \cite{2022AJ....163...25L}. We further refine these properties using  extinction-corrected infrared color excesses:  E($K_{\rm s}-$W2), E($K_{\rm s}-$W3) and E($K_{\rm s}-$W4), E($K_{\rm s}-$3.6), E($K_{\rm s}-$4.5) and E($K_{\rm s}-$5.8), E($K_{\rm s}-$8.0), and E($K_{\rm s}-$24). Among the 8,846 sources with estimates of stellar parameters ($A_{\rm V}$, SpT, and etc.), 95\% (8402/8846) of them have been detected with either the WISE or Spitzer missions. Figure.~\ref{Fig:EKIR} shows these extinction-corrected infrared color excesses as a function of spectral types. For each spectral type, we calculate the standard deviation ($\sigma_{0}$) of the excesses for the diskless sources initially identified in \cite{2022AJ....163...25L}. 

Sources meeting the following criterion are initially classified as exhibiting excess emission:
\begin{equation}
   \frac{Excess}{\sqrt{\sigma_{0}^2+\sigma_{Excess}^2}}\geq3
\end{equation}
\noindent 
where Excess denotes the extinction-corrected infrared color excess, $\sigma_{Excess}$  
is the uncertainty in the excess measurement, and $\sigma_{0}$ 
represents the standard deviation of excess values for diskless sources from \citet{2022AJ....163...25L}.

Initially, 1,450 sources were identified as candidates showing excess emission in at least one color band. We refined their disk classifications by visually comparing their SEDs to expected photospheric emission and inspecting WISE images to exclude false positives caused by photometric anomalies or contamination from nearby sources. A total of 295 sources were excluded from the excess-emission group. Among these excluded sources, 291 were originally classified as diskless in \citet{2022AJ....163...25L}.

The final sample includes 1,155 sources with infrared excess emission and 7,247 diskless sources. Among the excess-emission sources, 43 were initially labeled as diskless in \citet{2022AJ....163...25L} but exhibit infrared color excesses (77\%, or 33/43, of these were measured using Spitzer data). Conversely, 60 sources originally classified as disk-bearing in \citet{2022AJ....163...25L} failed to meet our excess-emission criteria and were reclassified as diskless.

The final disk classifications are summarized in Table~\ref{tabe_UpperSco}, which lists 7,247 diskless sources and 1,155 disk-bearing sources.

\section{Discussion}

\subsection{\newrev The Reconciliation of the Age Discrepancy}

It has been established that the age of a stellar population estimated using isochrones strongly depends on their spectral types \citep{2008ASPC..384..200H,2015ApJ...808...23H}. Leveraging the large population within the Sco-Cen complex, we can conduct a comprehensive study on this age dependency for four spectral type bins: F types, G types, K types, and M types. Our analysis reveals that the dependence of ages on spectral type is evident across all models considered and with different SPT-$T_{\rm eff}$ conversions.
Furthermore, there are clear age differences between the HRD-based method and the CMD-based method. These demonstrate that the choice of diagram and conversion method can significantly impact age estimates. 

Canonical evolutionary models typically ignore the effects of magnetic fields or cool spots on the appearance of young low-mass stars in the HRD. The Feiden and SPOTs models specifically account for these factors, which shifts the position of stars of the same mass to lower temperatures.

The median ages of K-type and M-type stars derived using the Feiden models are approximately 1.5 to 2 times older than those estimated by most canonical evolutionary models \citep[see also][e.g.]{2016A&A...593A..99F}, with the exception of PARCSEC. However, with the exception of the young Ophiuchus group, where the ages from both models are comparable, the median ages from the Feiden models still exhibit a dependence on spectral type for the US, UCL/LCC, and V1062 Sco groups. Specifically, the median ages for K-type stars are older than those for M-type stars in these groups.

Low-mass pre-main sequence (PMS) stars often exhibit cool spots, which significantly affect their observed properties. The extent of cool spot coverage varies depending on factors such as the star's age, rotation rate, and magnetic activity level \citep{2020ApJ...893...67M}. In extreme cases, like LkCa~4 in Taurus, the coverage can reach up to 80\% \citep{2017ApJ...836..200G}. For young PMS stars in Taurus (approximately 2.5\,Myr), the coverage ranges from 40\% to 85\%, with a median value of 67\% \citep{2024ApJ...967...45P}. This coverage decreases to 24.8\% in the Pleiades, at an age of approximately 125\,Myr, and further to 3\% in M67 at an age of approximately 4.3\,Gyr \citep{2022MNRAS.517.2165C}. The spot coverage roughly follows a $t^{-0.4}$ relationship \citep{2020ApJ...893...67M},  where $t$ is the age, leading to an estimated coverage of 34\%-45\% for sources aged 10-20~Myr.

\begin{figure*}
\begin{center}
\includegraphics[width=2\columnwidth]{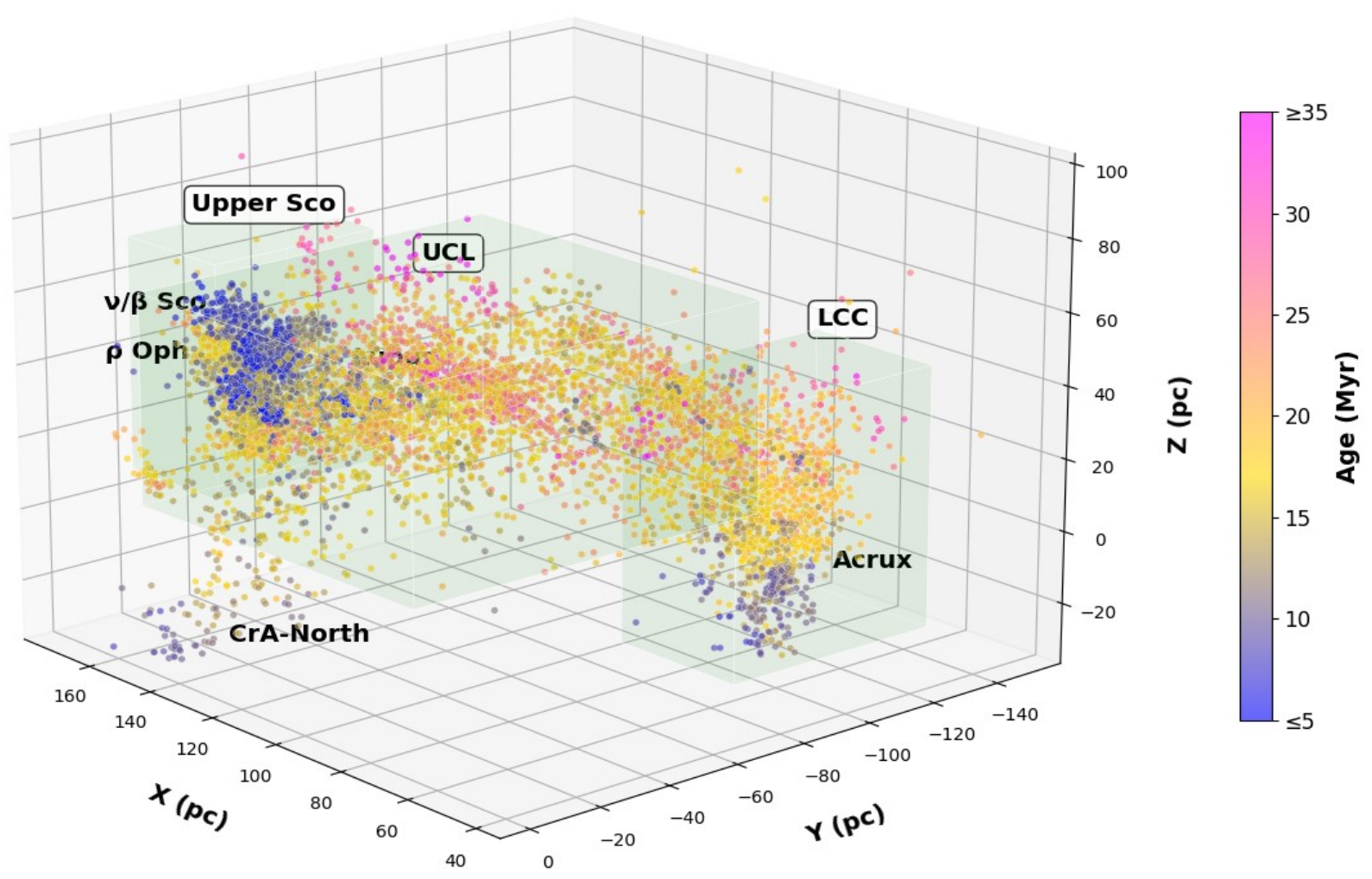}
\caption{Three-dimensional spatial distribution of young stars in the Sco-Cen complex, with stellar ages color-coded. Each star's age is smoothed by averaging with the ages of its 9 nearest neighbors. The Galactic XYZ coordinate system is adopted, with the Sun positioned at (0,0,0) and the Z=0 plane aligned parallel to the Galactic plane.
}\label{Fig:SFH}
\end{center}
\end{figure*}

\setcounter{table}{2}
\begin{table}
\renewcommand{\tabcolsep}{0.04cm}
{\newrev
\caption{\newrev Ages for the subgroups in Sco-cen complex derived in this work compared with those in the literature.}\label{Table:csamples_ages}
\begin{center}
\begin{tabular}{lcccccc}
\hline
\hline
Subgroups & Samples & Ages  & Models & Ref.  \\
          &  & [Myr]  & & & \\
\hline 
Upper~Sco &F-Type & 13 &Canonical & PM16\\
          &G-Type & 10  &Canonical & PM16  \\
          &K-M Type & 5  &Canonical & PM16  \\
          &G-M Type & 5  & Canonical & PZ99 \\
          &F-G Type & 12 &Canonical & this work\\
          &K-M Type & 12 &Spots & this work\\
\hline         
UCL       &B-Type & 16 & Canonical & MML02 \\ 
         &F-Type & 16 &Canonical & PM16\\
          &G-Type & 15  &Canonical & PM16  \\
          &K-M Type & 9  &Canonical & PM16  \\
          &F-G Type & 17 &Canonical & this work\\
          &K-M Type & 19 &Spots & this work\\          
\hline    
LCC       &B-Type & 17 &Canonical &MML02 \\
          &F-Type & 17 &Canonical & PM16\\
          &G-Type & 15  &Canonical & PM16  \\
          &K-M Type & 8  &Canonical & PM16  \\
          &F-G Type & 17 &Canonical & this work\\
          &K-M Type & 17 &Spots & this work\\          
          \hline
\end{tabular}
\end{center}
\tablecomments{MML02: \citet{2002AJ....124.1670M}; PM16:  \citet{2016MNRAS.461..794P}; PZ99: \citet{1999AJ....117.2381P}.
The ages listed for F-G stars in this work are derived from the Pisa model, while those for K-M stars are calculated using the SPOTS models with $f_{\rm spot}=$0.51 and the PM13 \& HH14 conversion.}}
\end{table}

\begin{figure*}
\begin{center}
\includegraphics[width=2\columnwidth]{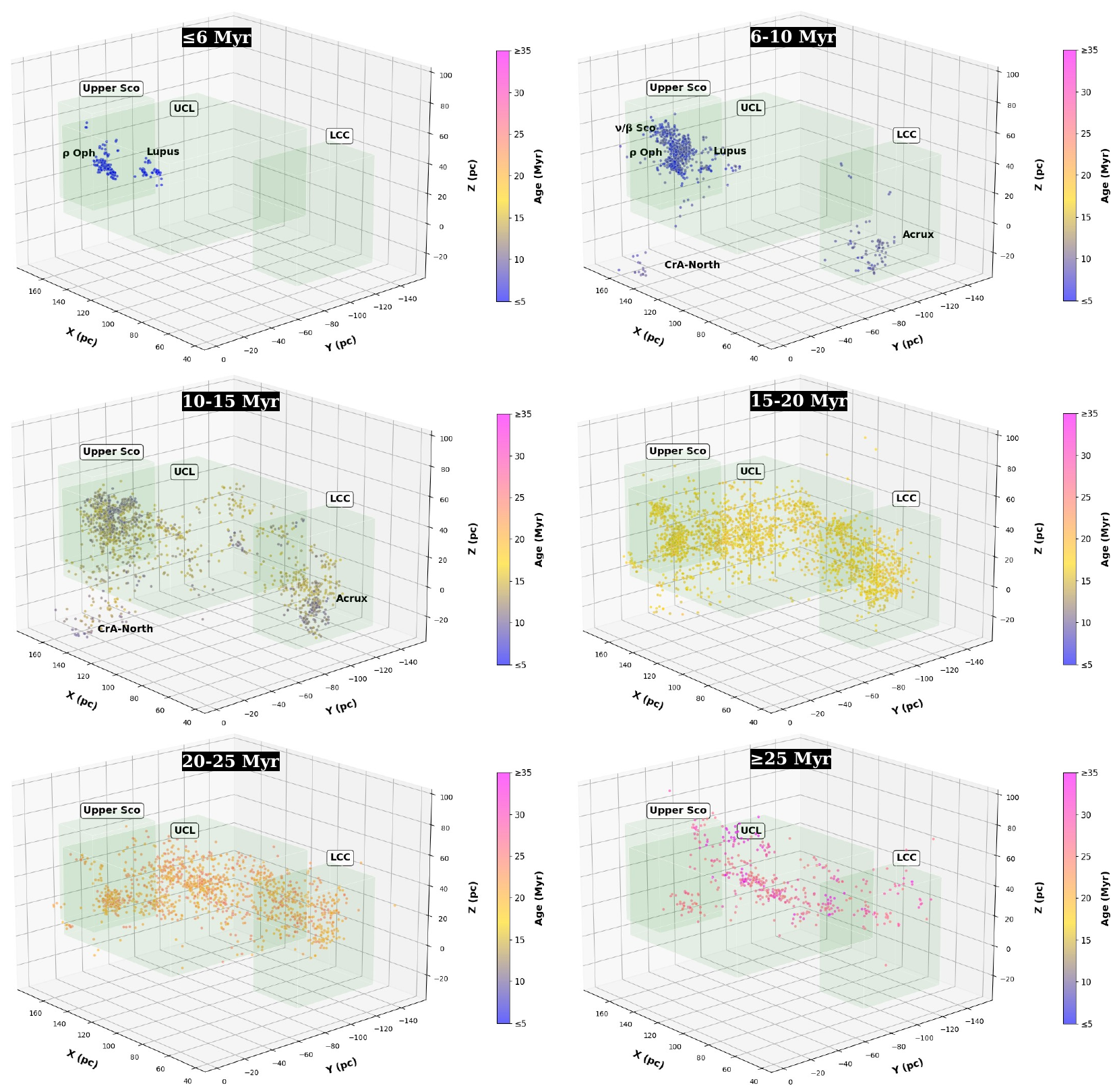}
\caption{Same as Figure~\ref{Fig:SFH} but show the sources within different age bins in each panel.
}\label{Fig:SFH_detail}
\end{center}
\end{figure*}

As shown in Figure~\ref{Fig:median_age}, the ages of K/M-type sources derived from the canonical models are significantly younger than those of F/G-type sources within the same group. This discrepancy arises because K/M-type stars are more influenced by magnetic activity. In contrast, F/G-type stars are less affected by these activities, resulting in similar age estimates across different canonical models. Because of this problem, the ages derived from canonical evolutionary models for F/G-type stars are likely more reliable. {\newrev To address the age discrepancy between K/M-type and F/G-type stars, the SPOTS models could be used to derive the ages for the K-M type stars. 
As shown in Figure~\ref{Fig:median_age}
\begin{itemize}
    \item Using the SPOTS models with $f_{\rm spot}=$0.51 and the PM13 \& HH14 conversion, or
    \item using $f_{\rm spot}=$0.34 and the KH95 \& L+03 conversion,
\end{itemize}
can reconcile the age discrepancies between different spectral type bins. 
}

{\newrev  To further validate our approach, Figure~\ref{Fig:agecom_pisa_Spots_US} presents age estimates for the US and UCL/LCC groups as a function of spectral type. Each group is divided into two catalogs: F/G-type sources (later than F5) and K0–M5-type sources. Ages for F/G-type sources are derived using PISA models, while those for K/M-type sources employ SPOTS models with  $f_{\rm spot}=$0.34 (using the KH95 \& L+03 conversion) and 0.51 (using the PM13 \& HH14 conversion). The median resulting ages of US-group sources are 12.3–13.0 Myr (F5–G), 13.2 Myr (K), and 11.1–11.6 Myr (M0–M5); UCL/LCC-group medians are 17.0–18.2 Myr (F5–G), 16.8–18.5 Myr (K), and 18.3–18.6 Myr (M0–M5), tightening age agreement between K/M- and F/G-type stars (see Figure~\ref{Fig:agecom_pisa_Spots_US}). For comparison, median Baraffe-model ages for K/M-type sources are 4–5 Myr and 5–8 Myr for the two groups (see age distributions in Figure~\ref{Fig:agecom_pisa_Spots_US}), respectively. }

{\newrev 
In Table~\ref{Table:csamples_ages}, we compare ages of the three groups (US, UCL and LCC) reported in the literature \citep{1999AJ....117.2381P,2002AJ....124.1670M,2016MNRAS.461..794P}. For F-G stars in this study, listed ages are derived from the Pisa model, whereas K-M star ages are calculated using the SPOTS models with $f_{\rm spot}=$0.51 and the PM13 \& HH14 conversion. Within each group, F-G star ages align with those of similarly spectrally typed stars reported in \cite{2016MNRAS.461..794P}. For the UCL and LCC groups, these ages are consistent with those derived from main-sequence turnoff ages of B-type stars \citep{2002AJ....124.1670M}. Additionally, K-M star ages from the SPOTS models here match the previously reported values and are approximately twice as old as samples of similar spectral types derived using canonical models in the literature \citep{1999AJ....117.2381P,2016MNRAS.461..794P}.
Overall, these comparisons confirm the consistency of our age determinations across spectral types with literature results and underscore the critical role of model selection in stellar age estimates \citep[for a comprehensive overview of young stellar ages, refer to the review by ][]{2014prpl.conf..219S}.}

Using the same setting for the SPOTS models, the median ages of the Ophiuchus are $\sim$6\,Myr.  Ongoing star formation in Ophiuchus should lead to a younger age, but the census of members from Gaia excludes the youngest stars in the region.

\subsection{Star formation history of the Sco-Cen Complex}

{\newrev 
In Figure \ref{Fig:SFH}, we show a 3D spatial distribution of stars (spectral types F5–M5) in the Sco-Cen complex, with stellar ages color-coded. For F/G-type stars, ages are computed using PISA models combined with the PM13 \& HH14 conversion; for K/M-type stars, SPOTS models (with $f_{\rm spot}=0.51$) and the same conversions are applied.
To mitigate noise from individual measurement uncertainties, each star’s age is further smoothed by averaging with the ages of its 9 nearest neighbors.} 

{\newrev Our age map (Figure~\ref{Fig:SFH}) is broadly consistent with the 2D age map of \cite{2016MNRAS.461..794P}, which used F/G/K-type stars for the same region, but reveals finer details to provide a more detailed view of the age distribution across the Sco-Cen complex.} \cite{2023A&A...678A..71R} decompose the Sco-Cen complex into 37 sub-groups and present a three-dimensional distribution of these groups, color-coded according to their ages. The detailed structures in our 3D age maps are consistent with the sub-group distributions identified by \cite{2023A&A...678A..71R}, reflecting similar age patterns and spatial organization, though their ages are estimated with the CMDs.

{\newrev To systematically investigate the spatial distribution of stellar populations in the Sco-Cen complex across distinct evolutionary stages, we 
stratify the stellar population within the Sco-Cen complex into six distinct age populations: $\leq$6\,Myr, $6-10$\,Myr, $10-15$\,Myr, $15-20$\,Myr, $20-25$\,Myr, and $25-35$\,Myr. This systematic division allows for a detailed investigation of how stellar spatial distributions evolve with age, with their three-dimensional configurations visually presented in Figure~\ref{Fig:SFH_detail}.
A notable trend emerges as we examine the data across these age bins: younger stellar populations ($\leq$10 Myr) exhibit tightly clustered spatial arrangements, characterized by compact groupings.  In contrast, older stellar populations ($\geq$15 Myr) display increasingly dispersed configurations, with stars spread over larger volumes. This transition from clustering to dispersion likely reflects the dynamical evolution of the Sco-Cen stellar population. Processes such as gravitational interactions between stars, gradual stellar migration due to velocity dispersions, and the dissipation of parental molecular cloud that once bound young clusters (via feedback from massive stars or stellar winds) may collectively drive this structural transformation.}

\begin{figure}
\begin{center}
\includegraphics[width=1\columnwidth]{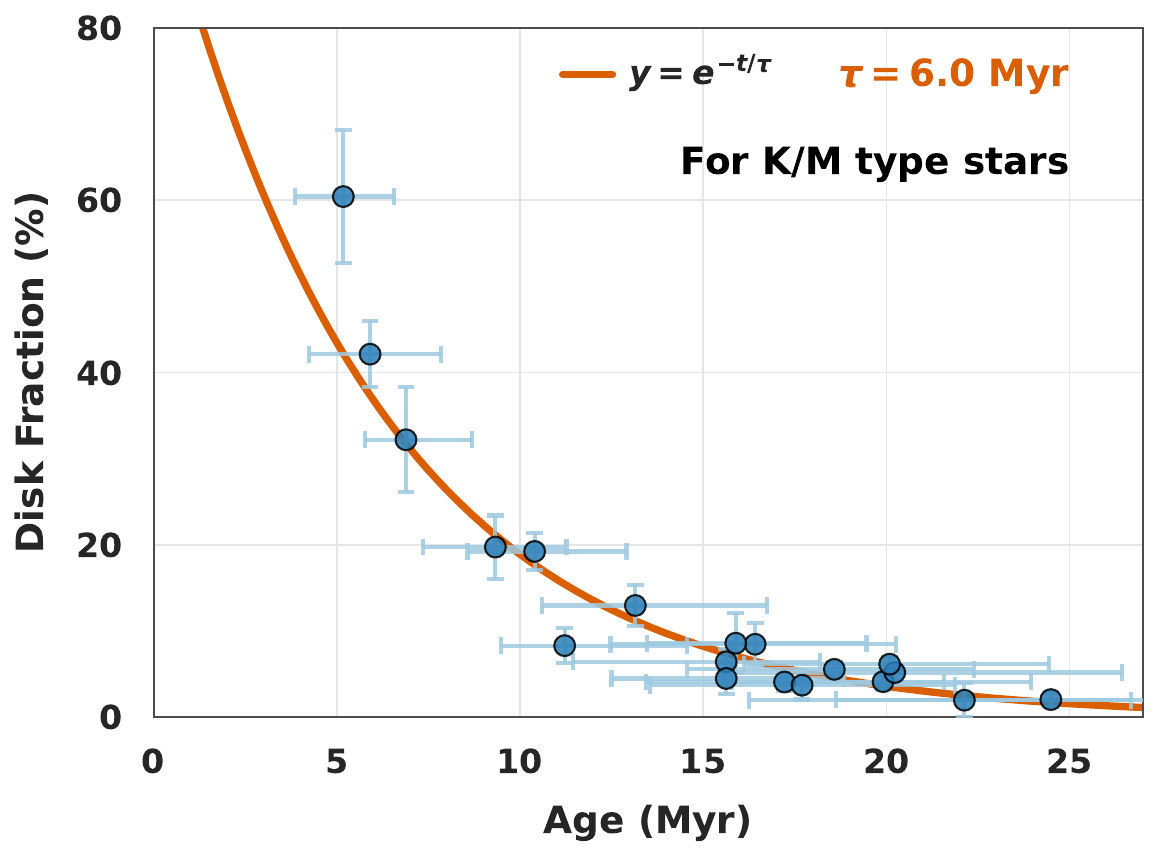}
\caption{Disk fractions of the groups in the Sco-Cen complex as a function of their ages for the K/M type sources.  
}\label{Fig:Disk}
\end{center}
\end{figure}

\begin{figure*}
\begin{center}
\includegraphics[width=2\columnwidth]{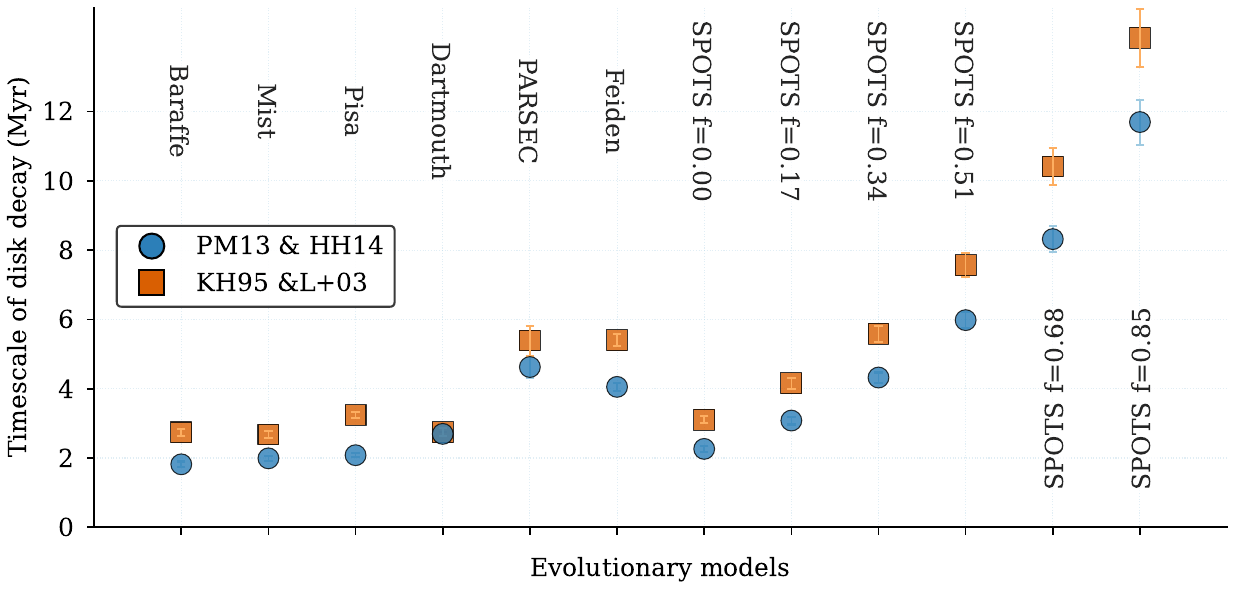}
\caption{The timescale of disk decay estimated with different evolutionary models. }\label{Fig:Disk_timescale}
\end{center}
\end{figure*}

\subsection{The disk lifetime in the Sco-cen complex}

As discussed above, the SPOTS models yield ages for K- and M-type stars that are consistent with those of F- and G-type stars within the same stellar groups. These revised age estimates are significantly older than those derived from canonical evolutionary models. Using the PM13 and HH14 calibration and SPOTS models with a spot coverage of $f_{\rm spot}=0.51$, we reassess the disk lifetimes of K- and M-type stars in our study region.

Sources are categorized into 37 distinct groups following the classification in \citet{2023A&A...677A..59R}. Of these groups, 21  contain over 50 usable members. For these individual groups, we calculated the median age and disk fraction of stars with spectral types between K0 to M5.
The disk fraction is defined as the proportion of sources exhibiting infrared excess emission in the WISE W2/W3 bands or Spitzer IRAC bands. Figure \ref{Fig:Disk} illustrates the relationship between disk fractions and ages for the 21 groups containing over 50 usable members. Assuming a disk decay model $DF=e^{-t/\tau}$, where $t$ is the group age, we fit the observed disk fraction versus age distribution. This yields a disk decay timescale of $\tau=6.0 \pm 0.2$\,Myr. This value is approximately double the previously reported values (2–3\,Myr) from similar studies using Spitzer IRAC bands \citep{2010A&A...510A..72F,2013A&A...549A..15F} or WISE bands shorter than W3 \citep{2014A&A...561A..54R}, which relied on canonical evolutionary models for age estimation.  Our longer disk lifetime is consistent with results from \citet{2013MNRAS.434..806B}, which adopted older cluster ages from isochrone fitting.

{\newrev To investigate the dependence of disk decay timescales on evolutionary models, we recalculated these timescales using distinct models and the two 
the two SPT-$T_{\rm eff}$ conversions. The results are presented in Figure~\ref{Fig:Disk_timescale}. For canonical models (Baraffe, MIST, Pisa, Dartmouth), the decay timescale ranges from 1.8 to 3.2 Myr, consistent with previous studies. In contrast, the PARSEC model yields a substantially longer timescale of 4.6–5.4 Myr. Feiden magnetic models project a decay timescale of 4.1–5.4 Myr, exceeding those of most canonical models. For SPOTS models, decay timescales increase progressively with spot coverage  $f_{\rm spot}$. Specifically, for cool spot coverages of 0.00, 0.17, 0.34, 0.51, 0.68, and 0.85, the timescales are 2.3, 3.1, 4.3, 6.0, 8.3, and 11.7 Myr (using the PM13 \& HH14 conversion) and 3.1, 4.2, 5.6, 7.6, 10.4, and 14.1 Myr (using the KH95 \& L+03 conversion), respectively.}

These results demonstrate that disk decay timescales for K/M-type stars are significantly prolonged when using the SPOTS models compared to canonical evolutionary frameworks, with the timescale increasing systematically as stellar cool spot coverage rises. 
 Our results underscore the critical influence of evolutionary model choice and magnetic activity on disk lifetime estimates.

\section{summary}
\label{sec:summary}

The Sco-Cen complex is one of the closest and most studied OB associations to the Sun. High-precision astrometric data from the Gaia mission has enhanced our understanding of Sco-Cen's structure and membership. Towards the new revealed member candidates, we conduct an extensive analysis of Gaia XP spectra. The main results  in this paper are summarized as follows: 

\begin{itemize}
    \item We construct Gaia XP spectral templates of young stars ranging from B2 to M8 spectral types. Using these templates, we classify the Gaia XP spectra of more than 7,800 potential members within the Sco-Cen complex. 
    
     \item By integrating the spectral classifications obtained in this study and from previous literature with archival photometric data, we estimate the V-band extinction and stellar luminosity for a total of 8,846 sources.
    \item The median ages of stellar groups within the Sco-Cen complex, when calculated using canonical evolutionary models, exhibit a dependence on spectral type, with discrepancies observed between earlier and later spectral types.
    \item There is a systematic shift between the ages derived from Hertzsprung-Russell Diagrams (HRDs) and those from Color-Magnitude Diagrams (CMDs), indicating that age estimates can vary depending on the method used. 
    \item A comparison of mass functions for the US and UCL/LCC groups  reveals consistency among canonical models, deviations in PARSEC models at lower masses, and systematically higher mass estimates from uncanonical models, especially SPOTS models with increased spot coverage.
    \item By applying the SPOTS models with $f_{\rm spot}$  of 0.34 (using the KH95 \&L+03 conversion) or 0.51 (using the PM13 \& HH14 conversion), we achieve consistent age estimates for K/M-type stars that align closely with those of F/G-type stars, revealing significantly older ages than previously estimated. This consistency allows for a re-evaluation of the star formation histories within the Sco-Cen complex, revealing evidence of substructure, and our revised age estimates yield an e-folding disk timescale of $\tau=6.0\pm0.2$ across the sub-groups within this region, approximately doubling previous estimates.  The disk lifetime may be considerably longer than previously thought.
    
\end{itemize}

\acknowledgments
This work is supported by the National Key R\&D Program of China with grant 2023YFA1608000 and the China Manned Space Program
with grant no. CMS-CSST-2025-A15.   GJH is supported by the general
grant 12173003 from the National Natural Science Foundation
of China and by National Key R\&D
program 2022YFA1603102 from the Ministry of
Science and Technology (MOST) of China.

\newpage
\setcounter{table}{0}
\rotate
\begin{deluxetable*}{rcccrccccccccccccccc}
\tablecaption{A List of the members with spectral types in Sco-Cen complex, as well as their stellar and disk properties (\textbf{the full table includes 8861 sources}) \label{tabe_UpperSco}}
\tablewidth{700pt}
\tabletypesize{\scriptsize}
\tablehead{
\colhead{ID}&\colhead{Gaia DR3}&\colhead{RA}& \colhead{DEC}   &\colhead{Dis} &\colhead{SpT1}  &\colhead{Ref$^{\alpha}$}&\colhead{SpT2}  &\colhead{$T_{\rm eff}$}&\colhead{Log${\rm Lum}$}&\colhead{$A_{\rm V}$}  &\colhead{W2/W3/W4$^{\beta}$} &\colhead{IR1/IR2/IR3/IR4/MIP1$^{\gamma}$} &\colhead{Disk}$^{\theta}$\\
\colhead{}&\colhead{}&\colhead{(J2016)} &\colhead{(J2016)}&\colhead{(pc)}&\colhead{(Lit)}&\colhead{}   &\colhead{(new)}&\colhead{(K)}        &\colhead{($L_{\odot}$)} &\colhead{(mag)}&\colhead{Excess} &\colhead{Excess}&\colhead{}}
\startdata
     42&    5369414461016305408 &  11 47 24.49&$-$49 53 03.2&113.8&G2.0&1&...&5870& 0.441& 0.0&N/N/N&U/U/U/U/Y&Y\\
     72&    5343610331876174336 &  11 53 07.93&$-$56 43 38.3&100.9&F3.0&1&F3.4&6635& 0.613& 0.0&N/N/Y&U/U/U/U/Y&Y\\
    101&    5341470132513014272 &  11 56 26.49&$-$58 49 17.0&107.2&F4.0&1&F3.1&6652& 0.593& 0.0&N/Y/Y&U/U/U/U/Y&Y\\
    134&    5335220267909593216 &  11 59 46.01&$-$61 01 13.4&118.4&K4.0&1&K3.5&4431&-0.223& 0.0&N/N/Y&U/U/U/U/U&Y\\
    139&    6075374902576831616 &  12 00 09.33&$-$57 07 02.1&103.2&F5.0&1&F5.6&6325& 0.512& 0.0&N/N/Y&U/U/U/U/Y&Y\\
    155&    5859187277060333184 &  12 02 37.58&$-$69 11 32.3&103.8&B9.0&1&B8.4&11715& 1.843& 0.0&N/Y/Y&U/U/U/U/Y&Y\\
    218&    6126469139186679808 &  12 08 04.81&$-$50 45 49.3&115.8&B9.0&1&B8.7&11252& 1.721& 0.0&N/N/Y&U/U/U/U/U&Y\\
    241&    6126242639795170688 &  12 09 02.20&$-$51 20 41.2&121.6&K3.0&1&K2.5&4660&-0.198& 0.0&N/N/Y&U/U/U/U/U&Y\\
    247&    6071348491003883136 &  12 09 38.71&$-$58 20 58.9&105.2&A3.0&1&A1.6&8978& 1.156& 0.3&N/N/Y&U/U/U/U/Y&Y\\
    272&    6072504073402829056 &  12 11 05.80&$-$56 24 05.1&113.4&A2.0&1&A2.7&8648& 1.151& 0.1&N/Y/Y&U/U/U/U/Y&Y
  \enddata
\tablecomments{$^{\alpha}$~References for the spectra type: 1. \cite{2022AJ....163...24L}, 2. \cite{2022AJ....163...26L}, 3. \cite{2021ApJ...908...49F}, 4. \cite{2017AJ....153..188F}, 5. \cite{Claes2024}; $^{\beta}$ and $^{\sigma}$  ~Infrared excess emission in individual WISE and Spitzer bands: Y for having excess emisson, N for no excess emission, U for unknown;  $^{\theta}$ Y for having disks, N for diskless, U for unknown}
\end{deluxetable*}


\vspace{5mm}

\newpage


\begin{thebibliography}{}
\expandafter\ifx\csname natexlab\endcsname\relax\def\natexlab#1{#1}\fi
\providecommand{\url}[1]{\href{#1}{#1}}

\bibitem[{{Allard}(2016)}]{2016sf2a.conf..223A}
{Allard}, F. 2016, in SF2A-2016: Proceedings of the Annual meeting of the
  French Society of Astronomy and Astrophysics, ed. C.~{Reyl{\'e}},
  J.~{Richard}, L.~{Cambr{\'e}sy}, M.~{Deleuil}, E.~{P{\'e}contal},
  L.~{Tresse}, \& I.~{Vauglin}, 223--227

\bibitem[{{Allard} {et~al.}(1997){Allard}, {Hauschildt}, {Alexander}, \&
  {Starrfield}}]{1997ARA&A..35..137A}
{Allard}, F., {Hauschildt}, P.~H., {Alexander}, D.~R., \& {Starrfield}, S.
  1997, \araa, 35, 137

\bibitem[{{Allard} {et~al.}(2011){Allard}, {Homeier}, \&
  {Freytag}}]{2011ASPC..448...91A}
{Allard}, F., {Homeier}, D., \& {Freytag}, B. 2011, in Astronomical Society of
  the Pacific Conference Series, Vol. 448, 16th Cambridge Workshop on Cool
  Stars, Stellar Systems, and the Sun, ed. C.~{Johns-Krull}, M.~K. {Browning},
  \& A.~A. {West}, 91

\bibitem[{{Allard} {et~al.}(2012){Allard}, {Homeier}, {Freytag}, \&
  {Sharp}}]{2012EAS....57....3A}
{Allard}, F., {Homeier}, D., {Freytag}, B., \& {Sharp}, C.~M. 2012, in EAS
  Publications Series, Vol.~57, EAS Publications Series, ed. C.~{Reyl{\'e}},
  C.~{Charbonnel}, \& M.~{Schultheis}, 3--43

\bibitem[{{Asplund} {et~al.}(2009){Asplund}, {Grevesse}, {Sauval}, \&
  {Scott}}]{2009ARA&A..47..481A}
{Asplund}, M., {Grevesse}, N., {Sauval}, A.~J., \& {Scott}, P. 2009, \araa, 47,
  481

\bibitem[{{Baraffe} {et~al.}(2015){Baraffe}, {Homeier}, {Allard}, \&
  {Chabrier}}]{2015A&A...577A..42B}
{Baraffe}, I., {Homeier}, D., {Allard}, F., \& {Chabrier}, G. 2015, \aap, 577,
  A42

\bibitem[{{Bell} {et~al.}(2013){Bell}, {Naylor}, {Mayne}, {Jeffries}, \&
  {Littlefair}}]{2013MNRAS.434..806B}
{Bell}, C. P.~M., {Naylor}, T., {Mayne}, N.~J., {Jeffries}, R.~D., \&
  {Littlefair}, S.~P. 2013, \mnras, 434, 806

\bibitem[{{Bressan} {et~al.}(2012){Bressan}, {Marigo}, {Girardi}, {Salasnich},
  {Dal Cero}, {Rubele}, \& {Nanni}}]{2012MNRAS.427..127B}
{Bressan}, A., {Marigo}, P., {Girardi}, L., {et~al.} 2012, \mnras, 427, 127

\bibitem[{{Brott} \& {Hauschildt}(2005)}]{2005ESASP.576..565B}
{Brott}, I., \& {Hauschildt}, P.~H. 2005, in ESA Special Publication, Vol. 576,
  The Three-Dimensional Universe with Gaia, ed. C.~{Turon}, K.~S. {O'Flaherty},
  \& M.~A.~C. {Perryman}, 565

\bibitem[{{Calvet} \& {Gullbring}(1998)}]{1998ApJ...509..802C}
{Calvet}, N., \& {Gullbring}, E. 1998, \apj, 509, 802

\bibitem[{{Cao} \& {Pinsonneault}(2022)}]{2022MNRAS.517.2165C}
{Cao}, L., \& {Pinsonneault}, M.~H. 2022, \mnras, 517, 2165

\bibitem[{{Cardelli} {et~al.}(1989){Cardelli}, {Clayton}, \&
  {Mathis}}]{1989ApJ...345..245C}
{Cardelli}, J.~A., {Clayton}, G.~C., \& {Mathis}, J.~S. 1989, \apj, 345, 245

\bibitem[{{Castelli} \& {Kurucz}(2003)}]{2003IAUS..210P.A20C}
{Castelli}, F., \& {Kurucz}, R.~L. 2003, in IAU Symposium, Vol. 210, Modelling
  of Stellar Atmospheres, ed. N.~{Piskunov}, W.~W. {Weiss}, \& D.~F. {Gray},
  A20

\bibitem[{{Chen} {et~al.}(2014){Chen}, {Girardi}, {Bressan}, {Marigo},
  {Barbieri}, \& {Kong}}]{2014MNRAS.444.2525C}
{Chen}, Y., {Girardi}, L., {Bressan}, A., {et~al.} 2014, \mnras, 444, 2525

\bibitem[{{Choi} {et~al.}(2016){Choi}, {Dotter}, {Conroy}, {Cantiello},
  {Paxton}, \& {Johnson}}]{2016ApJ...823..102C}
{Choi}, J., {Dotter}, A., {Conroy}, C., {et~al.} 2016, \apj, 823, 102

\bibitem[{{Claes} {et~al.}(2024{\natexlab{a}}){Claes}, {Campbell-White},
  {Manara}, {Frasca}, {Natta}, {Alcal{\'a}}, {Armeni}, {Fang}, {Lovell},
  {Stelzer}, {Venuti}, {Wyatt}, \& {Queitsch}}]{Claes2024}
{Claes}, R.~A.~B., {Campbell-White}, J., {Manara}, C.~F., {et~al.}
  2024{\natexlab{a}}, \aap, 690, A122

\bibitem[{{Claes} {et~al.}(2024{\natexlab{b}}){Claes}, {Campbell-White},
  {Manara}, {Frasca}, {Natta}, {Alcal{\'a}}, {Armeni}, {Fang}, {Lovell},
  {Stelzer}, {Venuti}, {Wyatt}, \& {Queitsch}}]{2024arXiv240711866C}
---. 2024{\natexlab{b}}, arXiv e-prints, arXiv:2407.11866

\bibitem[{{Cutri} {et~al.}(2013){Cutri}, {Wright}, {Conrow}, {Fowler},
  {Eisenhardt}, {Grillmair}, {Kirkpatrick}, {Masci}, {McCallon}, {Wheelock},
  {Fajardo-Acosta}, {Yan}, {Benford}, {Harbut}, {Jarrett}, {Lake}, {Leisawitz},
  {Ressler}, {Stanford}, {Tsai}, {Liu}, {Helou}, {Mainzer}, {Gettings},
  {Gonzalez}, {Hoffman}, {Marsh}, {Padgett}, {Skrutskie}, {Beck}, {Papin}, \&
  {Wittman}}]{2013wise.rept....1C}
{Cutri}, R.~M., {Wright}, E.~L., {Conrow}, T., {et~al.} 2013, {Explanatory
  Supplement to the AllWISE Data Release Products}, Explanatory Supplement to
  the AllWISE Data Release Products, by R. M. Cutri et al., ,

\bibitem[{{Damiani} {et~al.}(2019){Damiani}, {Prisinzano}, {Pillitteri},
  {Micela}, \& {Sciortino}}]{2019A&A...623A.112D}
{Damiani}, F., {Prisinzano}, L., {Pillitteri}, I., {Micela}, G., \&
  {Sciortino}, S. 2019, \aap, 623, A112

\bibitem[{{de Zeeuw} {et~al.}(1999){de Zeeuw}, {Hoogerwerf}, {de Bruijne},
  {Brown}, \& {Blaauw}}]{1999AJ....117..354D}
{de Zeeuw}, P.~T., {Hoogerwerf}, R., {de Bruijne}, J.~H.~J., {Brown}, A.~G.~A.,
  \& {Blaauw}, A. 1999, \aj, 117, 354

\bibitem[{{Delfini} {et~al.}(2025){Delfini}, {Vioque}, {Ribas}, \&
  {Hodgkin}}]{2025A&A...699A.145D}
{Delfini}, L., {Vioque}, M., {Ribas}, {\'A}., \& {Hodgkin}, S. 2025, \aap, 699,
  A145

\bibitem[{{Dotter} {et~al.}(2008){Dotter}, {Chaboyer}, {Jevremovi{\'c}},
  {Kostov}, {Baron}, \& {Ferguson}}]{2008ApJS..178...89D}
{Dotter}, A., {Chaboyer}, B., {Jevremovi{\'c}}, D., {et~al.} 2008, \apjs, 178,
  89

\bibitem[{{Edenhofer} {et~al.}(2024){Edenhofer}, {Zucker}, {Frank}, {Saydjari},
  {Speagle}, {Finkbeiner}, \& {En{\ss}lin}}]{2024A&A...685A..82E}
{Edenhofer}, G., {Zucker}, C., {Frank}, P., {et~al.} 2024, \aap, 685, A82

\bibitem[{{Esplin} \& {Luhman}(2022)}]{2022AJ....163...64E}
{Esplin}, T.~L., \& {Luhman}, K.~L. 2022, \aj, 163, 64

\bibitem[{{Fang} {et~al.}(2020){Fang}, {Hillenbrand}, {Kim}, {Findeisen},
  {Herczeg}, {Carpenter}, {Rebull}, \& {Wang}}]{2020ApJ...904..146F}
{Fang}, M., {Hillenbrand}, L.~A., {Kim}, J.~S., {et~al.} 2020, \apj, 904, 146

\bibitem[{{Fang} {et~al.}(2021){Fang}, {Kim}, {Pascucci}, \&
  {Apai}}]{2021ApJ...908...49F}
{Fang}, M., {Kim}, J.~S., {Pascucci}, I., \& {Apai}, D. 2021, \apj, 908, 49

\bibitem[{{Fang} {et~al.}(2013){Fang}, {van Boekel}, {Bouwman}, {Henning},
  {Lawson}, \& {Sicilia-Aguilar}}]{2013A&A...549A..15F}
{Fang}, M., {van Boekel}, R., {Bouwman}, J., {et~al.} 2013, \aap, 549, A15

\bibitem[{{Fang} {et~al.}(2009){Fang}, {van Boekel}, {Wang}, {Carmona},
  {Sicilia-Aguilar}, \& {Henning}}]{2009A&A...504..461F}
{Fang}, M., {van Boekel}, R., {Wang}, W., {et~al.} 2009, \aap, 504, 461

\bibitem[{{Fang} {et~al.}(2017{\natexlab{a}}){Fang}, {Kim}, {Pascucci}, {Apai},
  {Zhang}, {Sicilia-Aguilar}, {Alonso-Mart{\'\i}nez}, {Eiroa}, \&
  {Wang}}]{2017AJ....153..188F}
{Fang}, M., {Kim}, J.~S., {Pascucci}, I., {et~al.} 2017{\natexlab{a}}, \aj,
  153, 188

\bibitem[{{Fang} {et~al.}(2017{\natexlab{b}}){Fang}, {Herczeg}, \&
  {Rizzuto}}]{2017ApJ...842..123F}
{Fang}, Q., {Herczeg}, G.~J., \& {Rizzuto}, A. 2017{\natexlab{b}}, \apj, 842,
  123

\bibitem[{{Fedele} {et~al.}(2010){Fedele}, {van den Ancker}, {Henning},
  {Jayawardhana}, \& {Oliveira}}]{2010A&A...510A..72F}
{Fedele}, D., {van den Ancker}, M.~E., {Henning}, T., {Jayawardhana}, R., \&
  {Oliveira}, J.~M. 2010, \aap, 510, A72

\bibitem[{{Feiden}(2016)}]{2016A&A...593A..99F}
{Feiden}, G.~A. 2016, \aap, 593, A99

\bibitem[{{Fouesneau} {et~al.}(2023){Fouesneau}, {Fr{\'e}mat}, {Andrae},
  {Korn}, {Soubiran}, {Kordopatis}, {Vallenari}, {Heiter}, {Creevey}, {Sarro},
  {de Laverny}, {Lanzafame}, {Lobel}, {Sordo}, {Rybizki}, {Slezak},
  {{\'A}lvarez}, {Drimmel}, {Garabato}, {Delchambre}, {Bailer-Jones},
  {Hatzidimitriou}, {Lorca}, {Le Fustec}, {Pailler}, {Mary}, {Robin},
  {Utrilla}, {Abreu Aramburu}, {Bakker}, {Bellas-Velidis}, {Bijaoui}, {Blomme},
  {Bouret}, {Brouillet}, {Brugaletta}, {Burlacu}, {Carballo}, {Casamiquela},
  {Chaoul}, {Chiavassa}, {Contursi}, {Cooper}, {Dafonte}, {Demouchy},
  {Dharmawardena}, {Garc{\'\i}a-Lario}, {Garc{\'\i}a-Torres}, {Gomez},
  {Gonz{\'a}lez-Santamar{\'\i}a}, {Jean-Antoine Piccolo}, {Kontizas},
  {Lebreton}, {Licata}, {Lindstr{\o}m}, {Livanou}, {Magdaleno Romeo},
  {Manteiga}, {Marocco}, {Martayan}, {Marshall}, {Nicolas}, {Ordenovic},
  {Palicio}, {Pallas-Quintela}, {Pichon}, {Poggio}, {Recio-Blanco}, {Riclet},
  {Santove{\~n}a}, {Schultheis}, {Segol}, {Silvelo}, {Smart}, {S{\"u}veges},
  {Th{\'e}venin}, {Torralba Elipe}, {Ulla}, {van Dillen}, {Zhao}, \&
  {Zorec}}]{2023A&A...674A..28F}
{Fouesneau}, M., {Fr{\'e}mat}, Y., {Andrae}, R., {et~al.} 2023, \aap, 674, A28

\bibitem[{{Gaia Collaboration} {et~al.}(2023{\natexlab{a}}){Gaia
  Collaboration}, {Montegriffo}, {Bellazzini}, {De Angeli}, {Andrae},
  {Barstow}, {Bossini}, {Bragaglia}, {Burgess}, {Cacciari}, {Carrasco},
  {Chornay}, {Delchambre}, {Evans}, {Fouesneau}, {Fr{\'e}mat}, {Garabato},
  {Jordi}, {Manteiga}, {Massari}, {Palaversa}, {Pancino}, {Riello}, {Ruz
  Mieres}, {Sanna}, {Santove{\~n}a}, {Sordo}, {Vallenari}, {Walton}, {Brown},
  {Prusti}, {de Bruijne}, {Arenou}, {Babusiaux}, {Biermann}, {Creevey},
  {Ducourant}, {Eyer}, {Guerra}, {Hutton}, {Klioner}, {Lammers}, {Lindegren},
  {Luri}, {Mignard}, {Panem}, {Pourbaix}, {Randich}, {Sartoretti}, {Soubiran},
  {Tanga}, {Bailer-Jones}, {Bastian}, {Drimmel}, {Jansen}, {Katz}, {Lattanzi},
  {van Leeuwen}, {Bakker}, {Casta{\~n}eda}, {Fabricius}, {Galluccio},
  {Guerrier}, {Heiter}, {Masana}, {Messineo}, {Mowlavi}, {Nicolas},
  {Nienartowicz}, {Pailler}, {Panuzzo}, {Riclet}, {Roux}, {Seabroke},
  {Th{\'e}venin}, {Gracia-Abril}, {Portell}, {Teyssier}, {Altmann}, {Audard},
  {Bellas-Velidis}, {Benson}, {Berthier}, {Blomme}, {Busonero}, {Busso},
  {C{\'a}novas}, {Carry}, {Cellino}, {Cheek}, {Clementini}, {Damerdji},
  {Davidson}, {de Teodoro}, {Nu{\~n}ez Campos}, {Dell'Oro}, {Esquej},
  {Fern{\'a}ndez-Hern{\'a}ndez}, {Fraile}, {Garc{\'\i}a-Lario}, {Gosset},
  {Haigron}, {Halbwachs}, {Hambly}, {Harrison}, {Hern{\'a}ndez}, {Hestroffer},
  {Hodgkin}, {Holl}, {Jan{\ss}en}, {Jevardat de Fombelle}, {Jordan},
  {Krone-Martins}, {Lanzafame}, {L{\"o}ffler}, {Marchal}, {Marrese},
  {Moitinho}, {Muinonen}, {Osborne}, {Pauwels}, {Recio-Blanco}, {Reyl{\'e}},
  {Rimoldini}, {Roegiers}, {Rybizki}, {Sarro}, {Siopis}, {Smith}, {Sozzetti},
  {Utrilla}, {van Leeuwen}, {Abbas}, {{\'A}brah{\'a}m}, {Abreu Aramburu},
  {Aerts}, {Aguado}, {Ajaj}, {Aldea-Montero}, {Altavilla}, {{\'A}lvarez},
  {Alves}, {Anderson}, {Anglada Varela}, {Antoja}, {Baines}, {Baker},
  {Balaguer-N{\'u}{\~n}ez}, {Balbinot}, {Balog}, {Barache}, {Barbato},
  {Barros}, {Bartolom{\'e}}, {Bassilana}, {Bauchet}, {Becciani}, {Berihuete},
  {Bernet}, {Bertone}, {Bianchi}, {Binnenfeld}, {Blanco-Cuaresma}, {Boch},
  {Bombrun}, {Bouquillon}, {Bramante}, {Breedt}, {Bressan}, {Brouillet},
  {Brugaletta}, {Bucciarelli}, {Burlacu}, {Butkevich}, {Buzzi}, {Caffau},
  {Cancelliere}, {Cantat-Gaudin}, {Carballo}, {Carlucci}, {Carnerero},
  {Casamiquela}, {Castellani}, {Castro-Ginard}, {Chaoul}, {Charlot}, {Chemin},
  {Chiaramida}, {Chiavassa}, {Comoretto}, {Contursi}, {Cooper}, {Cornez},
  {Cowell}, {Crifo}, {Cropper}, {Crosta}, {Crowley}, {Dafonte}, {Dapergolas},
  {David}, {de Laverny}, {De Luise}, {De March}, {De Ridder}, {de Souza}, {de
  Torres}, {del Peloso}, {del Pozo}, {Delbo}, {Delgado}, {Delisle}, {Demouchy},
  {Dharmawardena}, {Diakite}, {Diener}, {Distefano}, {Dolding}, {Enke},
  {Fabre}, {Fabrizio}, {Faigler}, {Fedorets}, {Fernique}, {Figueras},
  {Fournier}, {Fouron}, {Fragkoudi}, {Gai}, {Garcia-Gutierrez},
  {Garcia-Reinaldos}, {Garc{\'\i}a-Torres}, {Garofalo}, {Gavel}, {Gavras},
  {Gerlach}, {Geyer}, {Giacobbe}, {Gilmore}, {Girona}, {Giuffrida}, {Gomel},
  {Gomez}, {Gonz{\'a}lez-N{\'u}{\~n}ez}, {Gonz{\'a}lez-Santamar{\'\i}a},
  {Gonz{\'a}lez-Vidal}, {Granvik}, {Guillout}, {Guiraud},
  {Guti{\'e}rrez-S{\'a}nchez}, {Guy}, {Hatzidimitriou}, {Hauser}, {Haywood},
  {Helmer}, {Helmi}, {Sarmiento}, {Hidalgo}, {H{\l}adczuk}, {Hobbs}, {Holland},
  {Huckle}, {Jardine}, {Jasniewicz}, {Jean-Antoine Piccolo},
  {Jim{\'e}nez-Arranz}, {Juaristi Campillo}, {Julbe}, {Karbevska}, {Kervella},
  {Khanna}, {Kordopatis}, {Korn}, {K{\'o}sp{\'a}l}, {Kostrzewa-Rutkowska},
  {Kruszy{\'n}ska}, {Kun}, {Laizeau}, {Lambert}, {Lanza}, {Lasne}, {Le
  Campion}, {Lebreton}, {Lebzelter}, {Leccia}, {Leclerc}, {Lecoeur-Taibi},
  {Liao}, {Licata}, {Lindstr{\'o}m}, {Lister}, {Livanou}, {Lobel}, {Lorca},
  {Loup}, {Madrero Pardo}, {Magdaleno Romeo}, {Managau}, {Mann}, {Marchant},
  {Marconi}, {Marcos}, {Marcos Santos}, {Mar{\'\i}n Pina}, {Marinoni},
  {Marocco}, {Marshall}, {Martin Polo}, {Mart{\'\i}n-Fleitas}, {Marton},
  {Mary}, {Masip}, {Mastrobuono-Battisti}, {Mazeh}, {McMillan}, {Messina},
  {Michalik}, {Millar}, {Mints}, {Molina}, {Molinaro}, {Moln{\'a}r}, {Monari},
  {Mongui{\'o}}, {Montero}, {Mor}, {Mora}, {Morbidelli}, {Morel}, {Morris},
  {Muraveva}, {Murphy}, {Musella}, {Nagy}, {Noval}, {Oca{\~n}a}, {Ogden},
  {Ordenovic}, {Osinde}, {Pagani}, {Pagano}, {Palicio}, {Pallas-Quintela},
  {Panahi}, {Payne-Wardenaar}, {Pe{\~n}alosa Esteller}, {Penttil{\"a}},
  {Pichon}, {Piersimoni}, {Pineau}, {Plachy}, {Plum}, {Poggio}, {Pr{\v{s}}a},
  {Pulone}, {Racero}, {Ragaini}, {Rainer}, {Raiteri}, {Ramos}, {Ramos-Lerate},
  {Re Fiorentin}, {Regibo}, {Richards}, {Rios Diaz}, {Ripepi}, {Riva}, {Rix},
  {Rixon}, {Robichon}, {Robin}, {Robin}, {Roelens}, {Rogues}, {Rohrbasser},
  {Romero-G{\'o}mez}, {Rowell}, {Royer}, {Rybicki}, {Sadowski}, {S{\'a}ez
  N{\'u}{\~n}ez}, {Sagrist{\`a} Sell{\'e}s}, {Sahlmann}, {Salguero}, {Samaras},
  {Sanchez Gimenez}, {Sarasso}, {Schultheis}, {Sciacca}, {Segol}, {Segovia},
  {S{\'e}gransan}, {Semeux}, {Shahaf}, {Siddiqui}, {Siebert}, {Siltala},
  {Silvelo}, {Slezak}, {Slezak}, {Smart}, {Snaith}, {Solano}, {Solitro},
  {Souami}, {Souchay}, {Spagna}, {Spina}, {Spoto}, {Steele},
  {Steidelm{\"u}ller}, {Stephenson}, {S{\"u}veges}, {Surdej}, {Szabados},
  {Szegedi-Elek}, {Taris}, {Taylor}, {Teixeira}, {Tolomei}, {Tonello}, {Torra},
  {Torra}, {Torralba Elipe}, {Trabucchi}, {Tsounis}, {Turon}, {Ulla}, {Unger},
  {Vaillant}, {van Dillen}, {van Reeven}, {Vanel}, {Vecchiato}, {Viala},
  {Vicente}, {Voutsinas}, {Wevers}, {Wyrzykowski}, {Yoldas}, {Yvard}, {Zhao},
  {Zorec}, {Zucker}, \& {Zwitter}}]{2023A&A...674A..33G}
{Gaia Collaboration}, {Montegriffo}, P., {Bellazzini}, M., {et~al.}
  2023{\natexlab{a}}, \aap, 674, A33

\bibitem[{{Gaia Collaboration} {et~al.}(2023{\natexlab{b}}){Gaia
  Collaboration}, {Vallenari}, {Brown}, {Prusti}, {de Bruijne}, {Arenou},
  {Babusiaux}, {Biermann}, {Creevey}, {Ducourant}, {Evans}, {Eyer}, {Guerra},
  {Hutton}, {Jordi}, {Klioner}, {Lammers}, {Lindegren}, {Luri}, {Mignard},
  {Panem}, {Pourbaix}, {Randich}, {Sartoretti}, {Soubiran}, {Tanga}, {Walton},
  {Bailer-Jones}, {Bastian}, {Drimmel}, {Jansen}, {Katz}, {Lattanzi}, {van
  Leeuwen}, {Bakker}, {Cacciari}, {Casta{\~n}eda}, {De Angeli}, {Fabricius},
  {Fouesneau}, {Fr{\'e}mat}, {Galluccio}, {Guerrier}, {Heiter}, {Masana},
  {Messineo}, {Mowlavi}, {Nicolas}, {Nienartowicz}, {Pailler}, {Panuzzo},
  {Riclet}, {Roux}, {Seabroke}, {Sordo}, {Th{\'e}venin}, {Gracia-Abril},
  {Portell}, {Teyssier}, {Altmann}, {Andrae}, {Audard}, {Bellas-Velidis},
  {Benson}, {Berthier}, {Blomme}, {Burgess}, {Busonero}, {Busso},
  {C{\'a}novas}, {Carry}, {Cellino}, {Cheek}, {Clementini}, {Damerdji},
  {Davidson}, {de Teodoro}, {Nu{\~n}ez Campos}, {Delchambre}, {Dell'Oro},
  {Esquej}, {Fern{\'a}ndez-Hern{\'a}ndez}, {Fraile}, {Garabato},
  {Garc{\'\i}a-Lario}, {Gosset}, {Haigron}, {Halbwachs}, {Hambly}, {Harrison},
  {Hern{\'a}ndez}, {Hestroffer}, {Hodgkin}, {Holl}, {Jan{\ss}en}, {Jevardat de
  Fombelle}, {Jordan}, {Krone-Martins}, {Lanzafame}, {L{\"o}ffler}, {Marchal},
  {Marrese}, {Moitinho}, {Muinonen}, {Osborne}, {Pancino}, {Pauwels},
  {Recio-Blanco}, {Reyl{\'e}}, {Riello}, {Rimoldini}, {Roegiers}, {Rybizki},
  {Sarro}, {Siopis}, {Smith}, {Sozzetti}, {Utrilla}, {van Leeuwen}, {Abbas},
  {{\'A}brah{\'a}m}, {Abreu Aramburu}, {Aerts}, {Aguado}, {Ajaj},
  {Aldea-Montero}, {Altavilla}, {{\'A}lvarez}, {Alves}, {Anders}, {Anderson},
  {Anglada Varela}, {Antoja}, {Baines}, {Baker}, {Balaguer-N{\'u}{\~n}ez},
  {Balbinot}, {Balog}, {Barache}, {Barbato}, {Barros}, {Barstow},
  {Bartolom{\'e}}, {Bassilana}, {Bauchet}, {Becciani}, {Bellazzini},
  {Berihuete}, {Bernet}, {Bertone}, {Bianchi}, {Binnenfeld}, {Blanco-Cuaresma},
  {Blazere}, {Boch}, {Bombrun}, {Bossini}, {Bouquillon}, {Bragaglia},
  {Bramante}, {Breedt}, {Bressan}, {Brouillet}, {Brugaletta}, {Bucciarelli},
  {Burlacu}, {Butkevich}, {Buzzi}, {Caffau}, {Cancelliere}, {Cantat-Gaudin},
  {Carballo}, {Carlucci}, {Carnerero}, {Carrasco}, {Casamiquela}, {Castellani},
  {Castro-Ginard}, {Chaoul}, {Charlot}, {Chemin}, {Chiaramida}, {Chiavassa},
  {Chornay}, {Comoretto}, {Contursi}, {Cooper}, {Cornez}, {Cowell}, {Crifo},
  {Cropper}, {Crosta}, {Crowley}, {Dafonte}, {Dapergolas}, {David}, {David},
  {de Laverny}, {De Luise}, {De March}, {De Ridder}, {de Souza}, {de Torres},
  {del Peloso}, {del Pozo}, {Delbo}, {Delgado}, {Delisle}, {Demouchy},
  {Dharmawardena}, {Di Matteo}, {Diakite}, {Diener}, {Distefano}, {Dolding},
  {Edvardsson}, {Enke}, {Fabre}, {Fabrizio}, {Faigler}, {Fedorets}, {Fernique},
  {Fienga}, {Figueras}, {Fournier}, {Fouron}, {Fragkoudi}, {Gai},
  {Garcia-Gutierrez}, {Garcia-Reinaldos}, {Garc{\'\i}a-Torres}, {Garofalo},
  {Gavel}, {Gavras}, {Gerlach}, {Geyer}, {Giacobbe}, {Gilmore}, {Girona},
  {Giuffrida}, {Gomel}, {Gomez}, {Gonz{\'a}lez-N{\'u}{\~n}ez},
  {Gonz{\'a}lez-Santamar{\'\i}a}, {Gonz{\'a}lez-Vidal}, {Granvik}, {Guillout},
  {Guiraud}, {Guti{\'e}rrez-S{\'a}nchez}, {Guy}, {Hatzidimitriou}, {Hauser},
  {Haywood}, {Helmer}, {Helmi}, {Sarmiento}, {Hidalgo}, {Hilger},
  {H{\l}adczuk}, {Hobbs}, {Holland}, {Huckle}, {Jardine}, {Jasniewicz},
  {Jean-Antoine Piccolo}, {Jim{\'e}nez-Arranz}, {Jorissen}, {Juaristi
  Campillo}, {Julbe}, {Karbevska}, {Kervella}, {Khanna}, {Kontizas},
  {Kordopatis}, {Korn}, {K{\'o}sp{\'a}l}, {Kostrzewa-Rutkowska},
  {Kruszy{\'n}ska}, {Kun}, {Laizeau}, {Lambert}, {Lanza}, {Lasne}, {Le
  Campion}, {Lebreton}, {Lebzelter}, {Leccia}, {Leclerc}, {Lecoeur-Taibi},
  {Liao}, {Licata}, {Lindstr{\o}m}, {Lister}, {Livanou}, {Lobel}, {Lorca},
  {Loup}, {Madrero Pardo}, {Magdaleno Romeo}, {Managau}, {Mann}, {Manteiga},
  {Marchant}, {Marconi}, {Marcos}, {Marcos Santos}, {Mar{\'\i}n Pina},
  {Marinoni}, {Marocco}, {Marshall}, {Martin Polo}, {Mart{\'\i}n-Fleitas},
  {Marton}, {Mary}, {Masip}, {Massari}, {Mastrobuono-Battisti}, {Mazeh},
  {McMillan}, {Messina}, {Michalik}, {Millar}, {Mints}, {Molina}, {Molinaro},
  {Moln{\'a}r}, {Monari}, {Mongui{\'o}}, {Montegriffo}, {Montero}, {Mor},
  {Mora}, {Morbidelli}, {Morel}, {Morris}, {Muraveva}, {Murphy}, {Musella},
  {Nagy}, {Noval}, {Oca{\~n}a}, {Ogden}, {Ordenovic}, {Osinde}, {Pagani},
  {Pagano}, {Palaversa}, {Palicio}, {Pallas-Quintela}, {Panahi},
  {Payne-Wardenaar}, {Pe{\~n}alosa Esteller}, {Penttil{\"a}}, {Pichon},
  {Piersimoni}, {Pineau}, {Plachy}, {Plum}, {Poggio}, {Pr{\v{s}}a}, {Pulone},
  {Racero}, {Ragaini}, {Rainer}, {Raiteri}, {Rambaux}, {Ramos}, {Ramos-Lerate},
  {Re Fiorentin}, {Regibo}, {Richards}, {Rios Diaz}, {Ripepi}, {Riva}, {Rix},
  {Rixon}, {Robichon}, {Robin}, {Robin}, {Roelens}, {Rogues}, {Rohrbasser},
  {Romero-G{\'o}mez}, {Rowell}, {Royer}, {Ruz Mieres}, {Rybicki}, {Sadowski},
  {S{\'a}ez N{\'u}{\~n}ez}, {Sagrist{\`a} Sell{\'e}s}, {Sahlmann}, {Salguero},
  {Samaras}, {Sanchez Gimenez}, {Sanna}, {Santove{\~n}a}, {Sarasso},
  {Schultheis}, {Sciacca}, {Segol}, {Segovia}, {S{\'e}gransan}, {Semeux},
  {Shahaf}, {Siddiqui}, {Siebert}, {Siltala}, {Silvelo}, {Slezak}, {Slezak},
  {Smart}, {Snaith}, {Solano}, {Solitro}, {Souami}, {Souchay}, {Spagna},
  {Spina}, {Spoto}, {Steele}, {Steidelm{\"u}ller}, {Stephenson}, {S{\"u}veges},
  {Surdej}, {Szabados}, {Szegedi-Elek}, {Taris}, {Taylor}, {Teixeira},
  {Tolomei}, {Tonello}, {Torra}, {Torra}, {Torralba Elipe}, {Trabucchi},
  {Tsounis}, {Turon}, {Ulla}, {Unger}, {Vaillant}, {van Dillen}, {van Reeven},
  {Vanel}, {Vecchiato}, {Viala}, {Vicente}, {Voutsinas}, {Weiler}, {Wevers},
  {Wyrzykowski}, {Yoldas}, {Yvard}, {Zhao}, {Zorec}, {Zucker}, \&
  {Zwitter}}]{2023A&A...674A...1G}
{Gaia Collaboration}, {Vallenari}, A., {Brown}, A.~G.~A., {et~al.}
  2023{\natexlab{b}}, \aap, 674, A1

\bibitem[{{Gully-Santiago} {et~al.}(2017){Gully-Santiago}, {Herczeg},
  {Czekala}, {Somers}, {Grankin}, {Covey}, {Donati}, {Alencar}, {Hussain},
  {Shappee}, {Mace}, {Lee}, {Holoien}, {Jose}, \& {Liu}}]{2017ApJ...836..200G}
{Gully-Santiago}, M.~A., {Herczeg}, G.~J., {Czekala}, I., {et~al.} 2017, \apj,
  836, 200

\bibitem[{{Hauschildt} {et~al.}(1999{\natexlab{a}}){Hauschildt}, {Allard}, \&
  {Baron}}]{1999ApJ...512..377H}
{Hauschildt}, P.~H., {Allard}, F., \& {Baron}, E. 1999{\natexlab{a}}, \apj,
  512, 377

\bibitem[{{Hauschildt} {et~al.}(1999{\natexlab{b}}){Hauschildt}, {Allard},
  {Ferguson}, {Baron}, \& {Alexander}}]{1999ApJ...525..871H}
{Hauschildt}, P.~H., {Allard}, F., {Ferguson}, J., {Baron}, E., \& {Alexander},
  D.~R. 1999{\natexlab{b}}, \apj, 525, 871

\bibitem[{{Herczeg} \& {Hillenbrand}(2014)}]{2014ApJ...786...97H}
{Herczeg}, G.~J., \& {Hillenbrand}, L.~A. 2014, \apj, 786, 97

\bibitem[{{Herczeg} \& {Hillenbrand}(2015)}]{2015ApJ...808...23H}
---. 2015, \apj, 808, 23

\bibitem[{{Hillenbrand} {et~al.}(2008){Hillenbrand}, {Bauermeister}, \&
  {White}}]{2008ASPC..384..200H}
{Hillenbrand}, L.~A., {Bauermeister}, A., \& {White}, R.~J. 2008, in
  Astronomical Society of the Pacific Conference Series, Vol. 384, 14th
  Cambridge Workshop on Cool Stars, Stellar Systems, and the Sun, ed. G.~{van
  Belle}, 200

\bibitem[{{Huang} {et~al.}(2024){Huang}, {Yuan}, {Xiang}, {Huang}, {Xiao},
  {Xu}, {Zhang}, {Yang}, {Niu}, \& {Gu}}]{2024ApJS..271...13H}
{Huang}, B., {Yuan}, H., {Xiang}, M., {et~al.} 2024, \apjs, 271, 13

\bibitem[{{Kenyon} \& {Hartmann}(1995)}]{1995ApJS..101..117K}
{Kenyon}, S.~J., \& {Hartmann}, L. 1995, \apjs, 101, 117

\bibitem[{{Kerr} {et~al.}(2023){Kerr}, {Kraus}, \&
  {Rizzuto}}]{2023ApJ...954..134K}
{Kerr}, R., {Kraus}, A.~L., \& {Rizzuto}, A.~C. 2023, \apj, 954, 134

\bibitem[{{Luhman}(2022{\natexlab{a}})}]{2022AJ....163...24L}
{Luhman}, K.~L. 2022{\natexlab{a}}, \aj, 163, 24

\bibitem[{{Luhman}(2022{\natexlab{b}})}]{2022AJ....163...25L}
---. 2022{\natexlab{b}}, \aj, 163, 25

\bibitem[{{Luhman}(2022{\natexlab{c}})}]{2022AJ....164..151L}
---. 2022{\natexlab{c}}, \aj, 164, 151

\bibitem[{{Luhman}(2023)}]{2023AJ....165...37L}
---. 2023, \aj, 165, 37

\bibitem[{{Luhman} \& {Esplin}(2020)}]{2020AJ....160...44L}
{Luhman}, K.~L., \& {Esplin}, T.~L. 2020, \aj, 160, 44

\bibitem[{{Luhman} \& {Esplin}(2022)}]{2022AJ....163...26L}
---. 2022, \aj, 163, 26

\bibitem[{{Luhman} {et~al.}(2003){Luhman}, {Stauffer}, {Muench}, {Rieke},
  {Lada}, {Bouvier}, \& {Lada}}]{2003ApJ...593.1093L}
{Luhman}, K.~L., {Stauffer}, J.~R., {Muench}, A.~A., {et~al.} 2003, \apj, 593,
  1093

\bibitem[{{Mamajek} {et~al.}(2002){Mamajek}, {Meyer}, \&
  {Liebert}}]{2002AJ....124.1670M}
{Mamajek}, E.~E., {Meyer}, M.~R., \& {Liebert}, J. 2002, \aj, 124, 1670

\bibitem[{{Manara} {et~al.}(2017){Manara}, {Frasca}, {Alcal{\'a}}, {Natta},
  {Stelzer}, \& {Testi}}]{2017A&A...605A..86M}
{Manara}, C.~F., {Frasca}, A., {Alcal{\'a}}, J.~M., {et~al.} 2017, \aap, 605,
  A86

\bibitem[{{Miret-Roig} {et~al.}(2024){Miret-Roig}, {Alves}, {Barrado},
  {Burkert}, {Ratzenb{\"o}ck}, \& {Konietzka}}]{2024NatAs...8..216M}
{Miret-Roig}, N., {Alves}, J., {Barrado}, D., {et~al.} 2024, Nature Astronomy,
  8, 216

\bibitem[{{Morris}(2020)}]{2020ApJ...893...67M}
{Morris}, B.~M. 2020, \apj, 893, 67

\bibitem[{{Muench} {et~al.}(2002){Muench}, {Lada}, {Lada}, \&
  {Alves}}]{2002ApJ...573..366M}
{Muench}, A.~A., {Lada}, E.~A., {Lada}, C.~J., \& {Alves}, J. 2002, \apj, 573,
  366

\bibitem[{{Onken} {et~al.}(2024){Onken}, {Wolf}, {Bessell}, {Chang}, {Luvaul},
  {Tonry}, {White}, \& {Da Costa}}]{2024arXiv240202015O}
{Onken}, C.~A., {Wolf}, C., {Bessell}, M.~S., {et~al.} 2024, arXiv e-prints,
  arXiv:2402.02015

\bibitem[{{Pang} {et~al.}(2021){Pang}, {Yu}, {Tang}, {Hong}, {Yuan},
  {Pasquato}, \& {Kouwenhoven}}]{2021ApJ...923...20P}
{Pang}, X., {Yu}, Z., {Tang}, S.-Y., {et~al.} 2021, \apj, 923, 20

\bibitem[{{Pecaut} \& {Mamajek}(2013)}]{2013ApJS..208....9P}
{Pecaut}, M.~J., \& {Mamajek}, E.~E. 2013, \apjs, 208, 9

\bibitem[{{Pecaut} \& {Mamajek}(2016)}]{2016MNRAS.461..794P}
---. 2016, \mnras, 461, 794

\bibitem[{{Pecaut} {et~al.}(2012){Pecaut}, {Mamajek}, \&
  {Bubar}}]{2012ApJ...746..154P}
{Pecaut}, M.~J., {Mamajek}, E.~E., \& {Bubar}, E.~J. 2012, \apj, 746, 154

\bibitem[{{P{\'e}rez Paolino} {et~al.}(2024){P{\'e}rez Paolino}, {Bary},
  {Hillenbrand}, \& {Markham}}]{2024ApJ...967...45P}
{P{\'e}rez Paolino}, F., {Bary}, J.~S., {Hillenbrand}, L.~A., \& {Markham}, M.
  2024, \apj, 967, 45

\bibitem[{{Preibisch} \& {Zinnecker}(1999)}]{1999AJ....117.2381P}
{Preibisch}, T., \& {Zinnecker}, H. 1999, \aj, 117, 2381

\bibitem[{{Ratzenb{\"o}ck} {et~al.}(2023{\natexlab{a}}){Ratzenb{\"o}ck},
  {Gro{\ss}schedl}, {M{\"o}ller}, {Alves}, {Bomze}, \&
  {Meingast}}]{2023A&A...677A..59R}
{Ratzenb{\"o}ck}, S., {Gro{\ss}schedl}, J.~E., {M{\"o}ller}, T., {et~al.}
  2023{\natexlab{a}}, \aap, 677, A59

\bibitem[{{Ratzenb{\"o}ck} {et~al.}(2023{\natexlab{b}}){Ratzenb{\"o}ck},
  {Gro{\ss}schedl}, {Alves}, {Miret-Roig}, {Bomze}, {Forbes}, {Goodman},
  {Hacar}, {Lin}, {Meingast}, {M{\"o}ller}, {Piecka}, {Posch}, {Rottensteiner},
  {Swiggum}, \& {Zucker}}]{2023A&A...678A..71R}
{Ratzenb{\"o}ck}, S., {Gro{\ss}schedl}, J.~E., {Alves}, J., {et~al.}
  2023{\natexlab{b}}, \aap, 678, A71

\bibitem[{{Ribas} {et~al.}(2014){Ribas}, {Mer{\'\i}n}, {Bouy}, \&
  {Maud}}]{2014A&A...561A..54R}
{Ribas}, {\'A}., {Mer{\'\i}n}, B., {Bouy}, H., \& {Maud}, L.~T. 2014, \aap,
  561, A54

\bibitem[{{Skrutskie} {et~al.}(2006){Skrutskie}, {Cutri}, {Stiening},
  {Weinberg}, {Schneider}, {Carpenter}, {Beichman}, {Capps}, {Chester},
  {Elias}, {Huchra}, {Liebert}, {Lonsdale}, {Monet}, {Price}, {Seitzer},
  {Jarrett}, {Kirkpatrick}, {Gizis}, {Howard}, {Evans}, {Fowler}, {Fullmer},
  {Hurt}, {Light}, {Kopan}, {Marsh}, {McCallon}, {Tam}, {Van Dyk}, \&
  {Wheelock}}]{2006AJ....131.1163S}
{Skrutskie}, M.~F., {Cutri}, R.~M., {Stiening}, R., {et~al.} 2006, \aj, 131,
  1163

\bibitem[{{Slesnick} {et~al.}(2006){Slesnick}, {Carpenter}, \&
  {Hillenbrand}}]{2006AJ....131.3016S}
{Slesnick}, C.~L., {Carpenter}, J.~M., \& {Hillenbrand}, L.~A. 2006, \aj, 131,
  3016

\bibitem[{{Slesnick} {et~al.}(2008){Slesnick}, {Hillenbrand}, \&
  {Carpenter}}]{2008ApJ...688..377S}
{Slesnick}, C.~L., {Hillenbrand}, L.~A., \& {Carpenter}, J.~M. 2008, \apj, 688,
  377

\bibitem[{{Soderblom} {et~al.}(2014){Soderblom}, {Hillenbrand}, {Jeffries},
  {Mamajek}, \& {Naylor}}]{2014prpl.conf..219S}
{Soderblom}, D.~R., {Hillenbrand}, L.~A., {Jeffries}, R.~D., {Mamajek}, E.~E.,
  \& {Naylor}, T. 2014, in Protostars and Planets VI, ed. H.~{Beuther}, R.~S.
  {Klessen}, C.~P. {Dullemond}, \& T.~{Henning}, 219--241

\bibitem[{{Somers} {et~al.}(2020){Somers}, {Cao}, \&
  {Pinsonneault}}]{2020ApJ...891...29S}
{Somers}, G., {Cao}, L., \& {Pinsonneault}, M.~H. 2020, \apj, 891, 29

\bibitem[{{Spina} {et~al.}(2017){Spina}, {Randich}, {Magrini}, {Jeffries},
  {Friel}, {Sacco}, {Pancino}, {Bonito}, {Bravi}, {Franciosini}, {Klutsch},
  {Montes}, {Gilmore}, {Vallenari}, {Bensby}, {Bragaglia}, {Flaccomio},
  {Koposov}, {Korn}, {Lanzafame}, {Smiljanic}, {Bayo}, {Carraro}, {Casey},
  {Costado}, {Damiani}, {Donati}, {Frasca}, {Hourihane}, {Jofr{\'e}}, {Lewis},
  {Lind}, {Monaco}, {Morbidelli}, {Prisinzano}, {Sousa}, {Worley}, \&
  {Zaggia}}]{2017A&A...601A..70S}
{Spina}, L., {Randich}, S., {Magrini}, L., {et~al.} 2017, \aap, 601, A70

\bibitem[{{Teplitz} {et~al.}(2012){Teplitz}, {Capak}, {Hanish}, {Brooke},
  {Colbert}, {Desai}, {Hoard}, {Howell}, {Laher}, \&
  {Noriega-Crespo}}]{2012AAS...21942806T}
{Teplitz}, H.~I., {Capak}, P., {Hanish}, D., {et~al.} 2012, in American
  Astronomical Society Meeting Abstracts, Vol. 219, American Astronomical
  Society Meeting Abstracts \#219, 428.06

\bibitem[{{Tian}(2020)}]{2020ApJ...904..196T}
{Tian}, H.-J. 2020, \apj, 904, 196

\bibitem[{{Tognelli} {et~al.}(2011){Tognelli}, {Prada Moroni}, \&
  {Degl'Innocenti}}]{2011A&A...533A.109T}
{Tognelli}, E., {Prada Moroni}, P.~G., \& {Degl'Innocenti}, S. 2011, \aap, 533,
  A109

\bibitem[{{Viana Almeida} {et~al.}(2009){Viana Almeida}, {Santos}, {Melo},
  {Ammler-von Eiff}, {Torres}, {Quast}, {Gameiro}, \&
  {Sterzik}}]{2009A&A...501..965V}
{Viana Almeida}, P., {Santos}, N.~C., {Melo}, C., {et~al.} 2009, \aap, 501, 965

\bibitem[{{White} \& {Hillenbrand}(2004)}]{2004ApJ...616..998W}
{White}, R.~J., \& {Hillenbrand}, L.~A. 2004, \apj, 616, 998

\bibitem[{{Zhang} {et~al.}(2023){Zhang}, {Green}, \&
  {Rix}}]{2023MNRAS.524.1855Z}
{Zhang}, X., {Green}, G.~M., \& {Rix}, H.-W. 2023, \mnras, 524, 1855

\end{thebibliography}



\appendix

\section{Construction of Gaia XP Spectral templates}
\label{Appen:template}
We construct the Gaia XP Spectral templates using the diskless young stars with known spectral types ranging from B2 to M9 in the literature \citep{2021ApJ...908...49F,2024arXiv240711866C,2017A&A...605A..86M,2016MNRAS.461..794P,2012ApJ...746..154P,2013ApJS..208....9P,2022AJ....163...64E,2020AJ....160...44L,2022AJ....163...24L,2023AJ....165...37L,2022AJ....164..151L}, in Taurus, Lupus, Chamaeleon~I,  $\sigma$~Ori, $\eta$~Cha, TW~Hya, Scorpius-Centaurus complex, and Corona~Australis. We visually inspected the Gaia XP spectra of individual stars and included the spectral data of 2487 sources in the template construction. 

We dereddened the XP spectra of individual sources using the extinction estimated with their spectral types and intrinsic colors corresponding to the spectral types \citep{2022AJ....163...24L}. The extinction law is from \citet{1989ApJ...345..245C}, adopting a total-to-selective extinction value of $R_{\rm V} = 3.1$. The spectral templates of individual spectral types are constructed with these dereddened XP spectra using median. 

We fit the the XP spectrum of each source using these templates by minimizing $\chi^2$ over two free parameters, spectral types and extinction. 
 We compared the newly derived spectral types with those in the literature and include the sources with the fitted spectral types similar to the ones in the literature. The criteria used here are $\Delta SPT$>3 subclass for those later than K0, $\Delta SPT$>3 subclass for those earlier than K0, $\Delta SPT\leq$2 subclass for those with SPT between K0 and M0, $\Delta SPT\leq$1 subclass for those with SPT between M0 and M6, and  $\Delta SPT\leq$1.5 subclass for those later than M6.  The sources passing the criteria are used for further template construction. 

We iterate
the above procedure five times to achieve the final set of spectral templates for Gaia XP spectra. For each spectral type, we require at least 5 XP spectra. For those spectral types without enough sources, we obtain the templates by the linearly interpolating the templates with the spectral types later and earlier than the ones. The interpolated templates include the ones for B3, B5, B7, A4, A6, G2, G4, and G9.  The final spectral templates cover a spectral range from B2 to M8. 

\begin{figure*}
\begin{center}
\includegraphics[width=1.0\columnwidth]{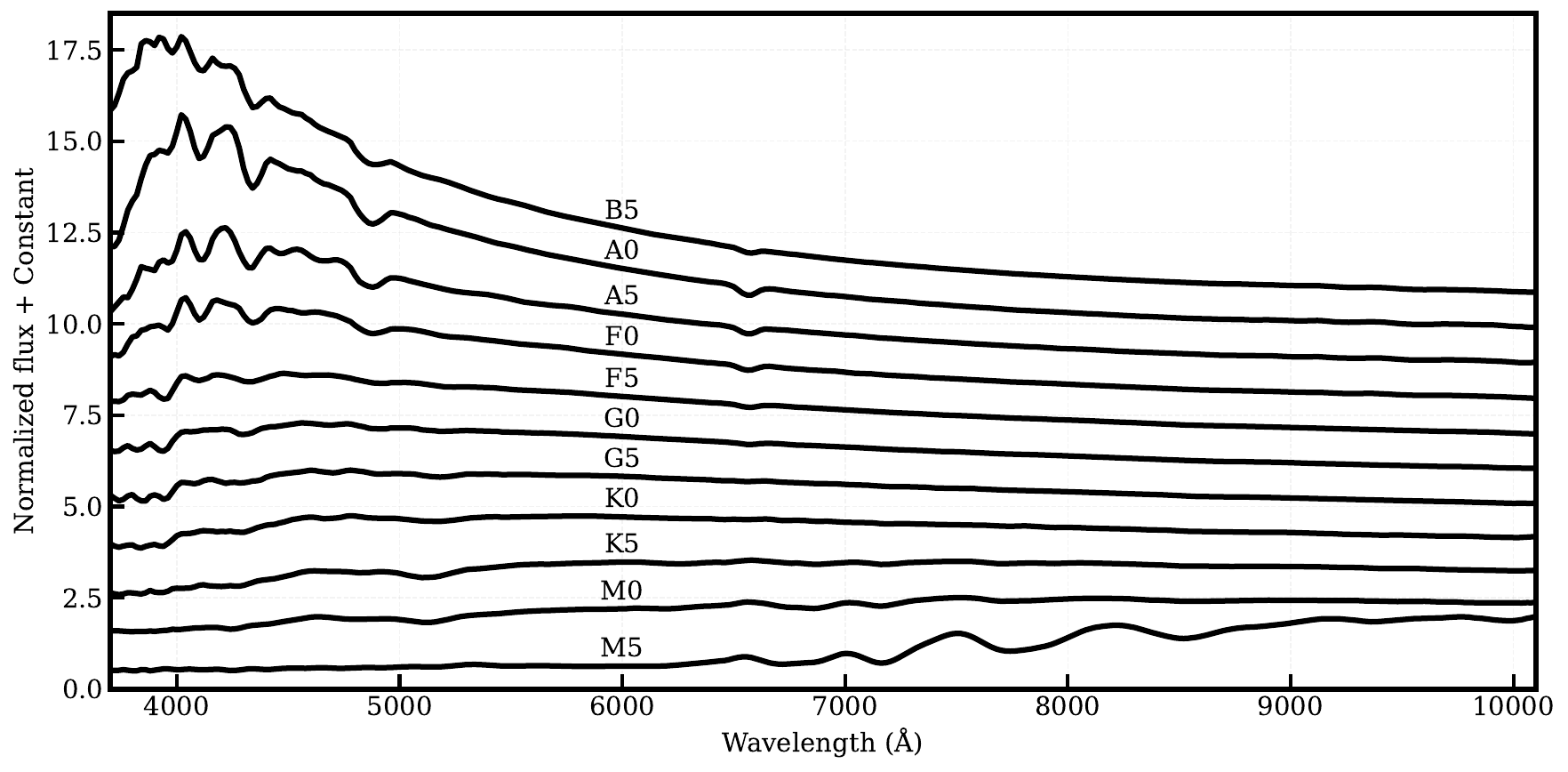}
\caption{
Example spectral templates of Gaia XP spectra for young stars constructed in this work.
 }\label{Fig:template}
\end{center}
\end{figure*}

\begin{figure*}
\begin{center}
\includegraphics[width=0.6\columnwidth]{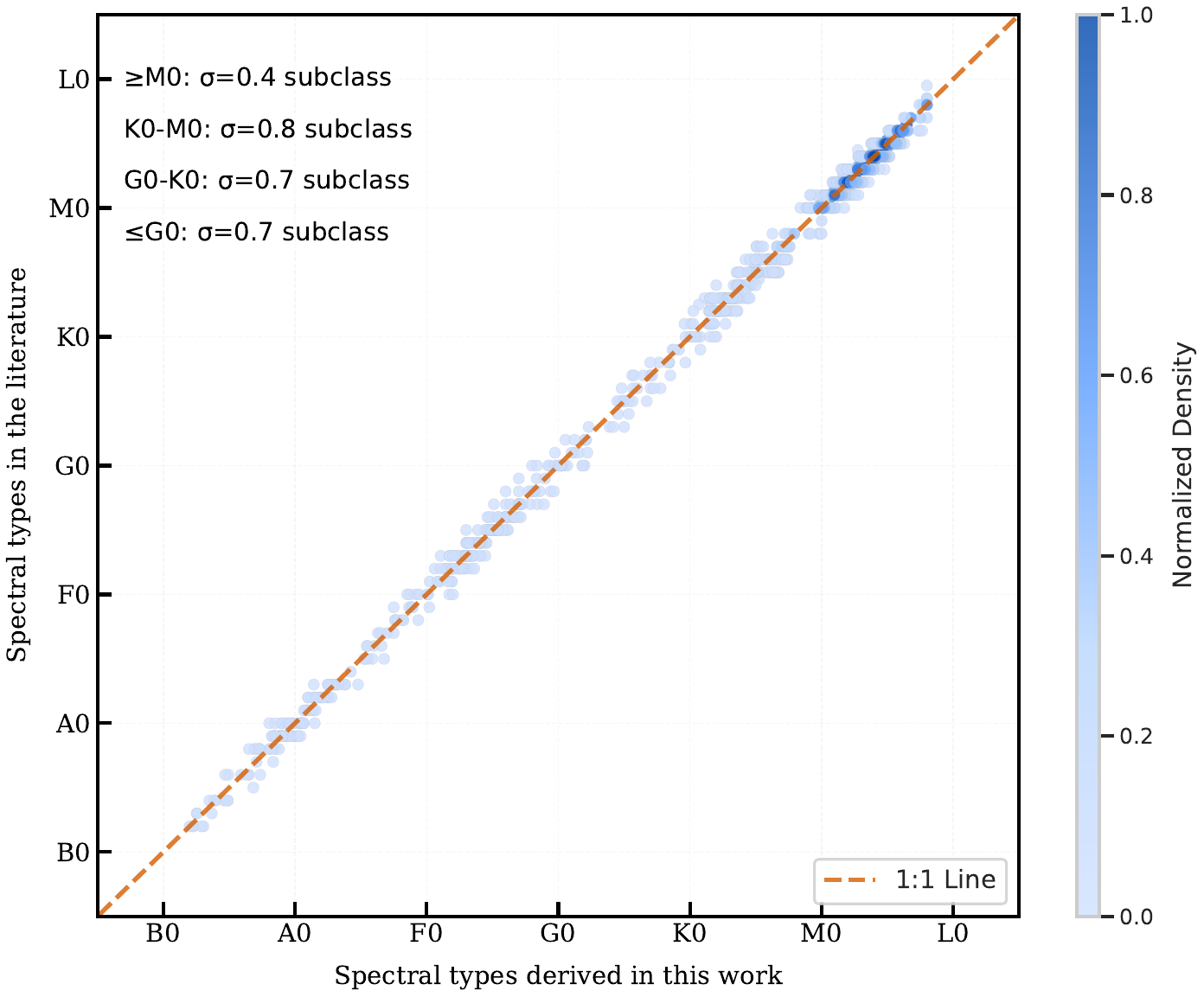}
\caption{
A comparison of spectral types derived in this work and in the literature for the sources used in the construction of spectral templates. The standard deviations on the SPT differences are 0.5, 0.8, 1.0, and 0.7 subclass for the sources with the SPTs later than M0, between K0 and M0, between G0 and K0, and earlier than G0, respectively. The color is represented by the normalized density.
  }\label{Fig:templatedif}
\end{center}
\end{figure*}

We show the example templates in Fig.~\ref{Fig:template}. In Fig.~\ref{Fig:templatedif}, we compare the two spectral types derived in this work and in the literature for the sources used for constructing the final templates. For the sources later than M0, the standard deviations on the SPT differences are 0.5
0.8, 1.0, and 0.7 subclass for the sources with the SPTs later than M0, between K0 and M0, between G0 and K0, and earlier than G0, respectively.

\section{Age Estimates Affected by Metallicity, Binarity, and Measurement Uncertainty }\label{Appen:age_effect}

{\newrev  Metallicity: Previous studies have reported solar metallicities for the US and LCC clusters \citep[-0.02$\pm$0.09dex and 0.02$\pm$0.05dex, respectively][]{2009A&A...501..965V}, and more generally all young clusters within 1 kpc all have roughly solar metallicity \citep{2017A&A...601A..70S}. The use of solar metallicity is therefore justified in this study.
For young K- and M-type stars modeled with PARSEC or MIST, a 0.2~dex increase in metallicity corresponds to an age offset of $+$0.13 to $+$0.18~dex, while a 0.2 dex decrease yields an offset of $-$0.13 to $-$0.17~dex.}

{\newrev Binary: Unresolved binaries can yield incorrect $T_{\rm eff}$ and elevated $Lum$, leading to inferred ages that are artificially young \citep{2017ApJ...842..123F}. To quantify this effect, we conducted a simple simulation for a population of model young K-M type stars at specific ages. Using the initial mass function (IMF) of the Trapezium cluster \citep{2002ApJ...573..366M}, we constructed a stellar population where $T_{\rm eff}$ 
and  $Lum$ for each star were derived from PARSEC and MIST evolutionary models at ages spanning 5–20\,Myr (in 3\,Myr intervals). Stars were randomly paired until 40\% of the sources were designated as binaries. For these binaries, $T_{\rm eff}$ was assumed to match that of the primary star, while $Lum$  was set as the sum of the primary and companion luminosities (consistent with \cite{2017ApJ...842..123F}). Under these assumptions, the resulting mean age of the stellar population with 40\% unresolved binaries is 0.07 dex younger than the input ages for both PARSEC and MIST models. In reality, the combined $T_{\rm eff}$ of a binary should be lower than that of the primary \citep[see discussion in][]{2014ApJ...786...97H}, further slightly reducing the inferred mean age.}

{\newrev Measurement Uncertainty: To address whether measurement uncertainties (in $T_{\rm eff}$, $Lum$, and parallax) are incorporated when determining the best-fitting age, we adopt a Monte Carlo approach to explicitly propagate these uncertainties. For each star, rather than using a single point estimate, we generate 1000 sets of $T_{\rm eff}$ and $Lum$ by sampling within their measurement errors (including contributions from spectral classification, SED fitting, and distance measurements). Ages are computed for each of these 1000 parameter combinations, and the median age across these realizations is taken as the final age estimate for the star. Using the PM13\& HH14 conversion, for F/G-type stars, applying the PISA model to the US and UCL/LCC groups yields  median logarithmic ages of 7.09 and 7.23, respectively; for K/M5-type stars, the SPOT ($f_{\rm spot}$=0.51) model produces median logarithmic ages of 7.07 (US) and 7.25 (UCL/LCC). The results are consistent with the ones without including the uncertainties. The conclusion is same for the ones using the KH95 \& L+03 conversion.
Notably, the typical age uncertainty for a single star, accounting for all parameters, is constrained to 0.01~dex for the F/G-type stars, and 0.05-0.08~dex for the K/M type stars. 

For the CMDs ($M_{\rm G}$ vs. BP-RP and $M_{\rm G}$ vs. G-RP), we follow the similar procedure as for the H–R diagram to include the uncertainties in photometry and distance for each source, and derive their ages and corresponding uncertainties. This yields the conclusions consistent with those obtained from the H–R diagrams.
}


\end{document}